\documentclass[ aps,  twocolumn, nofootinbib, showkeys, notitlepage, superscriptaddress]{revtex4-1}

\usepackage{graphicx}
\usepackage{dcolumn}
\usepackage{bm}
\usepackage{amsmath}
\usepackage{float}
\usepackage{multirow}
\usepackage{slashed}
\usepackage{xcolor}
\usepackage{physics}
\usepackage{multirow}
\usepackage{gensymb}
\usepackage[colorlinks=true, pdfstartview=FitV, bookmarks=true, bookmarksnumbered=true, breaklinks]{hyperref}
\usepackage{mathtools,braket}

\usepackage{lipsum}  
\usepackage{color}
\definecolor{blue}{rgb}{0.0, 0.0, 1.0}
\definecolor{red}{rgb}{1.0, 0.0, 0.0}
\definecolor{royalblue}{rgb}{0.0, 0.14, 0.4}
\hypersetup{linkcolor=red, citecolor=blue, urlcolor=blue}

\def\orcid#1{\kern .08em\href{https://orcid.org/#1}{\includegraphics[keepaspectratio,width=0.7em]{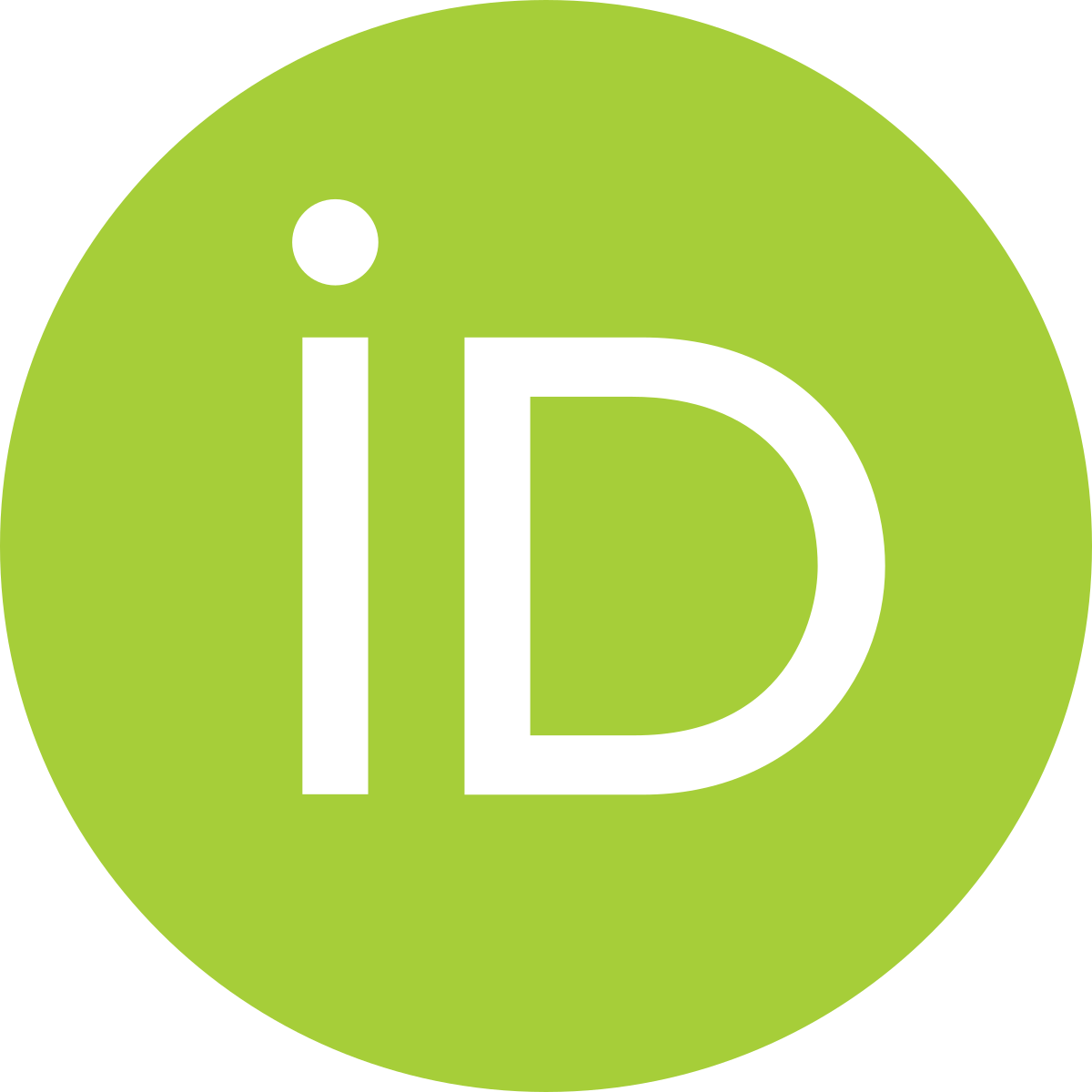}}}
\usepackage{calligra}
\DeclareMathAlphabet{\mathcalligra}{T1}{calligra}{m}{n}

\def\la{\langle}
\def\ra{\rangle}

\def\be{\begin{equation}}
\def\ee{\end{equation}}
\def\bea{\begin{eqnarray}}
\def\eea{\end{eqnarray}}

\begin{document}
\title{Pseudoscalar meson decay constants and distribution amplitudes up to the twist-4\\
in the light-front quark model}

\author{Ahmad Jafar Arifi\orcid{0000-0002-9530-8993}}
\email{ahmad.arifi@riken.jp}
\affiliation{Few-Body Systems in Physics Laboratory, RIKEN Nishina Center, Wako 351-0198, Japan}
\affiliation{Research Center for Nuclear Physics (RCNP), Osaka University, Ibaraki, Osaka 567-0047, Japan}
\affiliation{Asia Pacific Center for Theoretical Physics (APCTP), Pohang, Gyeongbuk 37673, Korea}

\author{Ho-Meoyng Choi\orcid{0000-0003-1604-7279}}
\email{homyoung@knu.ac.kr}
\affiliation{Department of Physics Education, Teachers College, Kyungpook National University, Daegu 41566, Korea}

\author{Chueng-Ryong Ji\orcid{0000-0002-3024-5186}}
\email{crji@ncsu.edu}
\affiliation{Department of Physics, North Carolina State University, Raleigh, NC 27695-8202, USA}

\date{\today}

\begin{abstract}
In the light-front quark model (LFQM) amenable to the simultaneous study of both the mass spectroscopy and the wave function related observables, we examine the decay constants and distribution amplitudes (DAs) up to the twist-4. The analysis of the heavy pseudoscalar mesons is carried out both in the $1S$ and $2S$ states. This investigation involves calculating the local and nonlocal matrix elements $\la 0|{\bar q}{\Gamma} q|P\ra$ using three distinct current operators ${\Gamma}=(\gamma^\mu\gamma_5, i\gamma_5,\sigma^{\mu\nu}\gamma_5)$. Considering a general reference frame where ${\bf P}_\perp\neq 0$ and investigating all available current components, we examine 
not only the frame-independence but also the component-independence of the decay constants. The explicit findings from our study provide the evidence for the equality of the three pseudoscalar meson decay constants obtained from the three distinct current operators $\Gamma$.
The notable agreement in decay constants is attained by imposing the Bakamjian-Thomas construction of the LFQM, namely the meson state is constructed by the noninteracting quark and antiquark representations while the interaction is added to the mass operator, 
which provides the self-consistency condition replacing the physical mass $M$ with the invariant mass $M_0$ for the noninteracting quark-antiquark representation of the meson state. In addition to obtaining the process-independent pseudoscalar meson decay constant, regardless of the choice of current operators $\Gamma$, 
we further demonstrate its explicit Lorentz and rotation invariance. In particular, we present 
the analysis conducted on the twist-4 DA derived from the minus component of the axial-vector current. Finally, we discuss the various twist DAs and their $\xi$-moments associated with the $1S$ and $2S$ heavy pseudoscalar mesons.
\end{abstract}

\maketitle

\section{Introduction}

The distribution amplitudes (DAs) of mesons are important non-perturbative ingredients in comprehending a range of 
the light-cone dominated processes that can be treated via collinear factorization~\cite{Lepage:1980fj,ER80,Chernyak:1983ej}, as they offer valuable insights into the nonperturbative makeup of hadrons and the distribution of partons in relation to their longitudinal momentum fractions within these particles.
The meson's DA is typically defined as a matrix element of a quark-antiquark bilocal light-cone operator between the vacuum and the meson state in the light-front dynamics (LFD)~\cite{BPP} which provides a natural separation of the meson's momentum into its longitudinal and transverse components. Thus, the LFD appears to be a practical and rigorous framework for computing the DAs of mesons categorizing them according to their increasing twist. 

While the leading-twist DA describes the longitudinal momentum distribution of valence quarks inside the meson providing a straightforward interpretation of the partonic structure of the meson, the higher-twist DAs are considerably more abundant as they take into account of various effects including the transverse motion of quarks or antiquarks, the higher Fock states involving extra gluons and/or quark-antiquark pairs, etc.~\cite{Ball:2006wn}. In the light-cone dominated hard processes, the leading-twist DAs give the dominant contributions and the higher twist contributions are suppressed by a power of the hard scale. As a result, the study of higher twist DAs has received less attention in comparison to the leading twist DAs in the analyses of the hard processes. However, with the higher statistical precision of experimental data expected from KEKII, LHC, JLAB, and the forthcoming Electron-Ion-Collider (EIC)~\cite{Accardi:2012qut, AbdulKhalek:2021gbh}, the relevance of the higher twist effects in hadron structure increases, accentuating the growing importance of further exploring higher twist contributions, e.g. in the formalism of TMD factorization~\cite{Braun:2022gzl}, in the wide-angle photoproduction of pions~\cite{PK18}, and in the nonleptonic $B$-meson decay~\cite{BBNS} etc. Thus, the quest to obtain essential nonperturbative insights into QCD has spurred numerous theoretical investigations aimed at calculating not only the leading-twist DA but also the higher-twist DAs using various nonperturbative techniques, such as the QCD sum rule~\cite{Ball:2006wn,Ball90,HWZ04,Agaev05,BMS06,MPS10}, the chiral-quark model derived from the instanton vacuum~\cite{PPRWG,NK06}, the Nambu-Jona-Lasinio (NJL) model~\cite{AB02,PR01}, the Dyson-Schwinger equation (DSE) approach~\cite{CRS13,SCCRSZ}, and the light-front quark model (LFQM)~\cite{Hwang10,CJ07,C13,C15,Choi:2017uos}.

In particular, the LFQM is the theoretical framework based on the LFD that has been highly successful in explaining simultaneously both the mass spectra and the wave function related observables including the 
electroweak properties of mesons~\cite{CJ99,CJ99B,HJZR,Dhiman,Arifi22,Jaus90,Jaus91,CCH97,Choi07}. 
In this model, mesons are treated as bound states of constituent quarks and antiquarks. The LFQM typically places 
the constituent quark and antiquark on their respective mass shells, and the spin-orbit (SO) wave function is obtained 
through the Melosh transformation~\cite{Mel1} which is independent of the interaction and is uniquely determined from the ordinary quantum state representation $J^{PC}$. For the construction of the more phenomenologically accessible LFQM, we applied the variational principle with the trial radial wave function, typically the Gaussian radial wave function, with the Melosh transformed spin-orbit wave functions for the on-mass shell constituent quark and antiquark to provide simultaneous analyses of both the mass spectra and the 
wave function related observables such as the decay constants, form factors, etc.~\cite{CJ99,CJ99B,HJZR,Dhiman,Arifi22,CCH97,Choi07}. Our LFQM follows the 
Bakamjian-Thomas (BT) construction~\cite{BT53,KP91}, where the meson state is constructed by the noninteracting quark and antiquark representations while the interaction is added to the mass operator via $M= M_0 + V_{Q{\bar Q}}$ applying the variational principle with typical Gaussian radial wave function as the trial wave function for the variational calculation. Due to the absence of manifest covariance, however, it is challenging to identify the LF zero-mode contributions in the phenomenological LFQM by itself. In particular, as the number of DAs proliferates with the higher twist, the computation of the higher twist DAs involves not only the good component (e.g. $J^+=J^0 + J^3$) of the current but also the bad component (e.g. $J^-=J^0 - J^3$) of the current to identify the proliferated number of DAs with the more number of current components. However, employing the bad component in the computation is often quite challenging due to the involvement of the light-front (LF) zero modes and/or the instantaneous contributions to restore the Lorentz and gauge invariance~\cite{Keister,deMelo:1997cb,CJ98,DSHwang,Jaus99,deMelo:02,BCJ01}. Thus, it is important to conduct a rigorous study of the higher twist DAs and address the challenges involved in order to gain the better understanding of the hadron structures.

To pin down the treacherous points involving the LF zero modes and the instantaneous contributions, one may utilize the manifestly covariant Bethe-Salpter (BS) field-theoretic model. Using the LF projection of the manifestly covariant BS model, one  can provide the corresponding LFQM with the multipole ansatz for the meson-quark vertex function~\cite{deMelo:1997cb,deMelo:02,BCJ01}. This type of LFQM is useful for providing the theoretical guidance on how to analyze the LF zero modes although the obtained results are in general semi-realistic. 
To account for the Lorentz structure of a hadronic matrix element, 
the light-like four-vector $\omega$ was introduced and the corresponding covariant approach was developed originally in Ref.\cite{Kar76}\footnote{In this formulation, the state vector is defined on a plane characterized by the invariant equation $\omega\cdot x=0$, where $\omega$ represents an arbitrary light-like four vector $\omega=(\omega_0, {\vec \omega})$ satisfying $\omega^2=0$. The special choice $\omega=(1,0,0,-1)$ corresponds to the LF or null plane $\omega\cdot x=x^+=x^0+x^3=0$.}. Subsequently,
the authors of Ref.~\cite{Carbonell98} developed a method for identifying and separating spurious contributions, enabling them to determine the physically meaningful contributions to the hadronic form factors and coupling constants that are independent of the choice of $\omega$.
By employing this covariant methodology as described in Refs.~\cite{Carbonell98, Kar76}, Jaus~\cite{Jaus99} employed a manifestly covariant BS model as a reference to devise a fundamentally distinct technique for addressing this issue.
In this approach, Jaus developed a way of identifying the LF zero-mode contributions by removing spurious contributions proportional to the light-like vector $\omega^\mu$ in the physical observables. 
As the $\omega$-dependent contributions violate the covariance, they may be eliminated by including the LF zero-mode contributions and at the same time restoring the covariance of the current matrix elements in the solvable BS model. Jaus identified the light-front zero-mode contribution corresponding it to the removal of the spurious $\omega$-dependence and then applied the LF zero-mode contributions identified in the BS model to the LFQM simply by replacing the multipole type vertex function in the BS model with the Gaussian radial wave function. 

However, two of us~\cite{C13} found that Jaus's prescription to identify the zero mode is only valid in the BS model with the multipole type vertex function but not in the LFQM with the Gaussian radial wave function as it causes a serious problem of impeding the self-consistency in the computation of physical observables, 
e.g. the decay constant of a $\rho$ meson gives different results for different polarization (longitudinal and transverse) states in the LFQM. This finding has also been confirmed by others~\cite{Chang:2018zjq,Chang22}, indicating that the LF zero-mode contributions do depend on the model wave functions. In Ref.~\cite{C13}, we then identified a specific matching condition for the first time between the manifestly covariant BS model and our LFQM, which we called the ``Type II" link (see Eq. (49) in~\cite{C13}). This unique matching condition ensures the self-consistency of the LFQM analysis. For example, it was demonstrated~\cite{C13} by using the ``Type II" link that the two $\rho$ meson decay constants obtained from the longitudinal and transverse polarizations exhibit the equivalence in the LFQM numerical results.
One of the key ingredients in the ``Type II" link is the replacement of the physical meson mass $M$ that appeared in the
integrand for the matrix element calculation with the invariant mass $M_0$, which is equivalent to 
imposing the on-mass shell condition of the constituent quark and antiquark in the LFQM.
Enforcing the on-mass shell condition for the constituents is tantamount to ensuring four-momentum conservation $P=p_1 +p_2$ at the meson-quark vertex, where the meson and quark (antiquark) momenta are denoted as $P$ and $p_{1(2)}$, respectively. 
Such a replacement ($M\to M_0$) is indeed consistent with the BT construction~\cite{BT53,KP91}, namely, the meson state is constructed by the noninteracting quark and antiquark representations while the interaction is added to the mass operator via $M= M_0 + V_{Q{\bar Q}}$.
This replacement can also be viewed as an effective way to include the LF zero-mode contributions and restore the Lorentz symmetry of the model, in particular the rotational symmetry, when compared to the covariant BS model~\cite{C13}.

Subsequent works using the same ``Type II" link~\cite{C13} have been made for the analyses of twist-2 and twist-3 DAs of the light pseudoscalar $(\pi, K)$ mesons~\cite{C15,Choi:2017uos} through the matrix elements $\la 0|{\bar q}(z){\Gamma} q(-z)|P\ra$ of the nonlocal operators ${\Gamma}=(\gamma^\mu\gamma_5, i\gamma_5,\sigma^{\mu\nu}\gamma_5)$, discussing the link between the chiral symmetry of QCD and the numerical results of the LFQM. In the very recent work~\cite{Arifi:2022qnd}, 
the decay constant for pseudoscalar meson with the axial-vector $(\gamma^\mu\gamma_5)$ current in the equal quark and antiquark mass case was investigated in the LFQM and the self-consistent result independent of the current components was obtained. The decay constant of the vector meson was also investigated for both longitudinal and transverse polarizations, obtaining a self-consistent result  independent of all possible combinations of the current components and the polarizations. In particular, in Ref~\cite{Arifi:2022qnd}, it was explicitly demonstrated that the decay constants obtained via the ``Type II" link between the BS model and the LFQM are precisely equivalent to those obtained directly in the LFQM, where the on-mass shell condition of the constituents is enforced.

In this work, we extend our previous LFQM analyses~\cite{C15,Choi:2017uos} for the decay constant and the DAs 
of the ground $1S$ state light pseudoscalar mesons through the matrix elements $\la 0|{\bar q}(z){\Gamma} q(-z)|P\ra$
of the operators ${\Gamma}=(\gamma^\mu\gamma_5, i\gamma_5,\sigma^{\mu\nu}\gamma_5)$ to include both $(1S, 2S)$ state heavy-light and heavy-heavy
pseudoscalar mesons~\cite{Arifi:2022qnd}. In particular, we shall explicitly show that 
the three pseudoscalar meson decay constants defined through the three different operators $\Gamma$ are all identical numerically in our LFQM constrained by the on-mass condition of the constituents. Namely, we obtain the process-independent
pseudoscalar meson decay constant regardless of the current operators $\Gamma$ used at the level of one-body current matrix element computation, as the independence of the current operators $\Gamma$ means the independence of the decay process.
The new two-particle twist-4 DA is also obtained from the minus component of the axial vector current ($\Gamma=\gamma^\mu\gamma_5$).
We also investigate the different helicity contributions to the decay constants defined through different operators $\Gamma$ 
and perform a quantitative analysis of each helicity component for different heavy-light and heavy-heavy pseudoscalar meson systems. 
For the numerical calculations, we present the results both for the ground state ($1S$) and the radially excited state ($2S$) of heavy pseudoscalar mesons, which were discussed in our recent work~\cite{Arifi22}. 
We then scrutinize the shape of the leading- and higher-twist DAs and their $\xi$-moments, where $\xi= 2x-1$ with the LF longitudinal momentum fraction $x$ of the constituent.
The ``Type II" link between the covariant BS model and the LFQM is further  discussed for the deeper understanding of the underlying physics involved.

The paper is organized as follows: In Section II, we describe the LFQM and the light-front wave functions of $1S$ and $2S$ pseudoscalar heavy meson. 
In Section III, we examine the pseudoscalar decay constants derived from the three distinct current operators $\Gamma=(\gamma^\mu\gamma_5, i\gamma_5, \sigma^{\mu\nu}\gamma_5)$ and establish the process independence and rotational invariance of the decay constants.
In Section IV, we discuss the DAs up to the twist-4 obtained from the three local and nonlocal current operators $\Gamma$ and their $\xi$-moments. 
Finally, we summarize our findings in Section V.
In the Appendix, the ``Type II" link between the manifestly covariant BS model and the LFQM is demonstrated for the completeness.

\section{Light-front quark model}
When applied to meson states reflecting the feature of BT construction~\cite{BT53,KP91}, 
the LFQM employs a noninteracting $q{\bar q}$ representation to describe the Fock state that 
is composed of the constituent quark ($q$) and antiquark (${\bar q})$ while the interactions are incorporated into the mass operator $M:= M_0 + V_{q{\bar q}}$ to ensure compliance with the group structure satisfying the Poincar\'e algebraic commutation relations. The interactions are then encoded 
in the light-front wave function (LFWF) $\Psi_{q{\bar q}}$, which satisfies the eigenvalue 
equation $H_{q\bar{q}}|\Psi_{q{\bar q}} \ra = (M_0 + V_{q\bar{q}})|\Psi_{q{\bar q}}\ra =M_{q\bar{q}}|\Psi\ra$
of our QCD-motivated effective Hamiltonian~\cite{CJ99,CJ99B,HJZR,Dhiman,Arifi22}.

Our LFQM for the $1S$ state~\cite{CJ99,CJ99B,HJZR,Dhiman,Arifi22}  and $2S$ state~\cite{Arifi22}  pseudoscalar and vector mesons is based on the central concept of using the radial wave function as a variational trial function for the QCD-motivated effective Hamiltonian $H_{q{\bar q}}$, which results in the determination 
of the mass eigenvalues $M_{q{\bar q}}$.
Once the values of the model parameters are determined by the variational analysis of 
the mass spectra, those determined model parameters are used to describe different observables including decay constants and electromagnetic and weak 
form factors etc.~\cite{CJ99,CJ99B,HJZR,Dhiman,Arifi22,CJ07,C21}

For the self-consistent analysis of the decay constants and the higher twist DAs for the $(1S, 2S)$ state heavy pseudoscalar mesons
performed in this work, 
we provide a brief overview of the LFWFs for $1S$ and $2S$ state heavy pseudoscalar mesons presented in Ref.~\cite{Arifi22}
focusing on the important aspects of LFWFs constrained by the on-mass shell condition of the constituents.

The four-momentum $P$ of the meson in terms of the LF components is defined as
$P=(P^+, P^-, {\bf P}_\perp)$, where $P^+=P^0 + P^3$ and
$P^-=P^0 -P^3$ are the LF longitudinal momentum and the LF energy, respectively, and ${\bf P}_\perp=(P^1, P^2)$
are the transverse momenta.  Here, we take the metric convention as $P^2 = P^+P^- - {\bf P}^2_\perp$.
The meson state $|{\rm M}(P, J, J_z)\ra$ of momentum $P$ and spin state
$(J, J_z)$ can then be constructed as follows~\cite{CCH97,Cheng04,Cheng:2004ew}

%%%%%%%%%5
\begin{eqnarray}\label{eq:1}
\ket{\rm M}
&=& \int \left[ {\rm d}^3{\bar p}_1 \right] \left[ {\rm d}^3{\bar p}_2 \right]  2(2\pi)^3 \delta^3 \left({\bar P}-{\bar p}_1-{\bar p}_2 \right) 
\nonumber\\ && \times \mbox{} 
\sum_{\lambda_1,\lambda_2} \Psi_{\lambda_1 \lambda_2}^{JJ_z}(x, {\bf k}_\perp)
\ket{q_{\lambda_1}(p_1) \bar{q}_{\lambda_2}(p_2) },
\quad
\end{eqnarray}
where $p^\mu_i$ and $\lambda_i$ are the on-mass shell ($p^2_i=m^2_i$) momenta and the helicities of the
constituent quark ($i=1$) and antiquark $(i=2)$, respectively,
with the LF three momentum defined by ${\bar p}=(p^+,{\bf p}_\perp)$ 
and $\left[ {\rm d}^3{\bar p} \right] \equiv {\rm d}p^+ {\rm d}^2\mathbf{p}_{\perp}/(16\pi^3)$.
The LF internal relative variables $(x, {\bf k}_\perp)$ are
defined by $x_i=p^+_i/P^+$ and ${\bf k}_{i\perp} = {\bf p}_{i\perp} - x_i {\bf P}_\perp$, where  
$\sum_i x_i=1$ and $\sum_i{\bf k}_{i\perp}=0$ and we set $x=x_1$ and ${\bf k}_\perp ={\bf k}_{1\perp}$.
This meson state satisfies the following normalization
\bea\label{norm1}
 &&\la{{\rm M}(P', J', J'_z)}|{\rm M}(P, J, J_z)\ra 
\nonumber\\
 && = 2 (2\pi)^3 P^+ \delta^3({\bar P}^\prime -{\bar P})\delta_{J' J}\delta_{J^\prime_z J_z}.
\eea

In momentum space, the LFWF $\Psi_{q{\bar q}}(x, {\bf k}_\perp)$ of a meson can be decomposed as
\begin{eqnarray}\label{eq:5}
		\Psi^{JJ_z}_{\lambda_1\lambda_2} (x, \mathbf{k}_{\bot}) = \Phi(x, \mathbf{k}_\bot)
		\  \mathcal{R}^{JJ_z}_{\lambda_1\lambda_2}(x, \mathbf{k}_\bot),
\end{eqnarray}
where $\Phi(x, \mathbf{k}_\bot)$ is the radial wave function that was used as our trial function for the mass spectroscopic analysis~\cite{CJ99,CJ99B,HJZR,Dhiman,Arifi22}
and $\mathcal{R}^{JJ_z}_{\lambda_1\lambda_2 }$ 
is the SO wave function obtained by the interaction-independent Melosh transformation~\cite{Mel1}
for the corresponding meson quantum number $J^{PC}$.

We should note that one crucial aspect of the LF formulation for a bound state, as depicted in Eq.~\eqref{eq:1}, 
is the frame-independence of the LFWF~\cite{BPP}. In other words, the hadron's internal variables $(x, {\bf k}_\perp)$ 
of the wave function remain unaffected by boosts to any physical $(P^+, {\bf P}_\perp)$ frame, which is not the case in the instant formulation.
We shall explicitly show this boost invariance of the decay constant computed in general ${\bf P}_\perp\neq 0$ frame.

The SO wave function for a pseudoscalar meson obtained from the interaction-independent Melosh transformation
can be written as the covariant form~\cite{Jaus90,Jaus91} consistent with the BT construction, which is given by
\begin{eqnarray}\label{eq:spinorbit}
	\mathcal{R}^{00}_{\lambda_1\lambda_2} 
	&=& \frac{\bar{u}_{\lambda_1}^{}(p_1) 
	\gamma_5
	v_{\lambda_2}^{}(p_2)}
	{\sqrt{2} {\tilde M_0}},\\
 &=& \frac{1}{\sqrt{2}\sqrt{\mathcal{A}^2 + \mathbf{k}_\perp^2}} 
\begin{pmatrix}
	-k^L 		& \mathcal{A}\\
		-\mathcal{A} & -k^R\\
\end{pmatrix}, 
\end{eqnarray}
where ${\tilde M_0}^2=M_0^2 - (m_1 -m_2)^2= (\mathcal{A}^2 + \mathbf{k}_\perp^2)/x(1-x)$,
$k^{R(L)}=k^1 \pm ik^2$ and $\mathcal{A}= (1-x)m_1 + x m_2$. The boost-invariant meson mass squared is
given by
\begin{eqnarray}
M_0^2 = \frac{\mathbf{k}_{\bot}^2 + m_1^2}{x}  + \frac{\mathbf{k}_{\bot}^2 + m_2^2}{1-x}.
\end{eqnarray}
Note that  the SO wave function satisfies the
unitary condition, $\sum_{\lambda's}{\cal R}^\dagger{\cal R}=1$.
It is worth noting that the LFWF $\Psi$ depends on the interaction-independent invariant mass $M_0$ that 
follows the BT construction as the meson is constructed in the noninteracting representation. 

For the $1S$ and $2S$ state radial wave functions $\Phi_{ns}$ of Eq.~(\ref{eq:5}), 
we allow the mixing between the two lowest order
harmonic oscillator (HO) wave functions $(\phi_{1S}, \phi_{2S})$ by writing~\cite{Arifi22}
\begin{eqnarray}
	\begin{pmatrix} \Phi_{1S} \\ \ \Phi_{2S}  \end{pmatrix} = 
	\begin{pmatrix} \cos\theta & \sin \theta  \\ -\sin \theta & \cos \theta \end{pmatrix} 
	\begin{pmatrix} {\phi}_{1S}  \\ {\phi}_{2S} \end{pmatrix},
\end{eqnarray}
where 
\begin{eqnarray}\label{HO1S2S}
	\phi_{1S} ({\vec k}) &=& \frac{4\pi^{3/4}}{\beta^{3/2}} e^{-{\vec k}^2/ 2\beta^2},
	\nonumber\\
	\phi_{2S} (\vec{k}) &=& \frac{4\pi^{3/4}}{\sqrt{6}\beta^{7/2}} \left( 2 {\vec k}^2 - 3\beta^2 \right)  e^{-{\vec k}^2/ 2\beta^2}.
\end{eqnarray}
Here, ${\vec k}=(k_z, \mathbf{k}_\perp)$ is the three momentum and 
$\beta$ represents a parameter that serves as the variational parameter in our mass spectroscopic analysis~\cite{Arifi22}.
The rotationally invariant HO wave functions $\phi_{nS}(\vec{k})$ in Eq.~\eqref{HO1S2S} satisfy 
\be\label{norm2}
\int\; \frac{{\rm d}^3{\vec k}}{2(2\pi)^3}\;\abs{ \phi_{nS}({\vec k}) }^2 =1.
\ee
Sustaining this rotationally invariant property of the wave function, 
one can transform the normalization of $\phi_{nS} ({\vec k})$ to that of $\phi_{nS} (x, {\bf k}_\perp)$
via the variable transformation $(k_z, \mathbf{k}_\perp) \to (x,\mathbf{k}_\perp)$ as follows
\be\label{eq:10}
 \int_0^1  {\rm d}x \int \frac{{\rm d}^2 \mathbf{k}_\bot}{2(2\pi)^3}  \abs{ \phi_{nS}(x, \mathbf{k}_\bot) }^2 =1.
\ee 
We note that the wave functions $\phi_{nS}(x, \mathbf{k}_\bot)$ include the Jacobian factor $\partial k_z/\partial x$ as
\begin{eqnarray}\label{HO1S2SJac}
    \phi_{nS}(x,\mathbf{k}_\perp) = \sqrt{\frac{\partial k_z}{\partial x}} \phi_{nS}(\vec{k})
\end{eqnarray}
because of the variable transformation $(k_z, \mathbf{k}_\perp) \to (x,\mathbf{k}_\perp)$ and
$k_z (=k^3)$ and $x$ are related by~\cite{CCH97}
\be\label{kz}
x = \frac{ E_1 - k_z}{E_1 + E_2}, \;  1 - x = \frac{E_2 + k_z}{E_1 + E_2},
\ee
where $E_i = \sqrt{m^2_i + {\vec k}^2}$. We then have $M_0 = E_1 + E_2$ and 
\begin{eqnarray}\label{eq:kz}
k_z = \left(x - \frac{1}{2}\right) M_0 + \frac{ m^2_1 - m^2_2}{2M_0}.
\end{eqnarray}
The Jacobian factor is then given by
\be\label{Jacob1}
\frac{\partial k_z}{\partial x} = \frac{E_1 E_2}{x (1-x) M_0},
\ee
or in terms of $(x, {\bf k})$
\be\label{Jacob2}
\frac{\partial k_z}{\partial x} = \frac{M_0}{4x(1-x)}\left[ 1 - \frac{(m_1^2 - m_2^2)^2}{M_0^4} \right].
\ee
It should be noted that the total LFWF $\Psi$ given by Eq.~\eqref{eq:5} meets the same normalization given by
Eq.~\eqref{eq:10}.
This is due to the meson state $|{\rm M}(P, J, J_z)\ra$ fulfilling the condition of  Eq.~\eqref{norm1},
and the SO wave function adhering to
the unitary condition. Especially, the inclusion of the Jacobian factor in defining $\phi_{nS}(x,\mathbf{k}_\perp)$ is the key aspect
to retaining the rotational invariance of the model wave function and obtaining the self-consistent, i.e. current-component and boost invariant, results
of physical observables. The quantitative effects of the Jacobian factor on the decay constants, twist-2 DAs, and electromagnetic form factors of
mesons were also discussed in Ref.~\cite{CJM0}.

%%% Table I

\section{Decay constants}\label{sec:DC}
For the decay constants and the leading-and higher-twist DAs of pseudoscalar mesons, one may obtain them from the 
matrix elements $\la 0|{\bar q}(z){\Gamma} q(-z)|P\ra$
of the the following three possible nonlocal operators ${\Gamma}=(\gamma^\mu\gamma_5, i\gamma_5,\sigma^{\mu\nu}\gamma_5)$,
where $z^\mu$ is the light-like vector ($z^2=0$). For the calculation of the decay constant $f_{\rm P}$, while $f_{\rm P}$ can be defined 
by using a local operator with axial-vector $(\Gamma_{\rm A}=\gamma^\mu \gamma_5)$ and pseudoscalar $(\Gamma_{\rm P}=i \gamma_5)$ current as~\cite{Ball:2006wn,Ball90} 
\begin{eqnarray}\label{eq:nonlocal1}
	\bra{0} \bar{q}(0) \gamma^\mu \gamma_5 q(0) \ket{{\rm P}(P)} &=& i f_{\rm P} P^\mu, \\
    \bra{0} \bar{q}(0) i\gamma_5 q(0) \ket{{\rm P}(P)} &=& f_{\rm P} \mu_M, \label{eq:nonlocal2}
\end{eqnarray} 
it can also be computed by utilizing the nonlocal matrix element
in the case of the pseudotensor current $(\Gamma_{\rm T}=\sigma^{\mu\nu}\gamma_5 )$
as defined by the subsequent equation~\cite{Ball90}:
\begin{eqnarray} \label{eq:tensor}   
\bra{0} \bar{q}(z)\sigma^{\mu\nu}\gamma_5 q(-z)\ket{{\rm P}(P)} 
    = -\frac{i}{3} f_{\rm P} \left(1 - \rho_+ \right) \mu_M\nonumber\\
 \times (P^\mu z^\nu - P^\nu z^\mu)  \int_0^1 {\rm d}x\ {\rm e}^{i\zeta P\cdot z}  \psi_{3;{\rm P}}(x),\quad \quad 
\end{eqnarray}
where $\sigma^{\mu\nu} = \frac{i}{2}[\gamma^\mu,\gamma^\nu]$, $\mu_M=M^2/(m_1+m_2)$, $\rho_+ = (m_1+m_2)^2/M^2$, and
$\zeta = 2x-1$. 
In this definition of Eq.~\eqref{eq:tensor}, the two-particle twist 3 DA $\psi_{3;{\rm P}}(x)$ is normalized to unity $\int^1_0 {\rm d}x \psi_{3;{\rm P}}(x) = 1$.
As a reference, in Ref.~\cite{Ball:2006wn}, defining the matrix element $\bra{0} \bar{q}(z)\sigma^{\mu\nu}\gamma_5 q(-z)\ket{{\rm P}(P)}$,
the authors removed the term $(1 - \rho_+)$ on the right-hand side (RHS) of Eq.~\eqref{eq:tensor} by normalizing $\psi_{3;{\rm P}}(x)$ in such a way that
$\int^1_0 {\rm d}x\; \psi_{3;{\rm P}}(x) = 1 - \rho_+$. In the previous LFQM analysis~\cite{Choi:2017uos} for $\psi_{3;{\rm P}}(x)$, two of us used the
definition of Ref.~\cite{Ball:2006wn} rather than Eq.~\eqref{eq:tensor}. 
However, in this study, we opt to use Eq.~\eqref{eq:tensor} as we observe that this definition yields the same decay constant as those obtained 
from Eqs.~\eqref{eq:nonlocal1} and~\eqref{eq:nonlocal2} in which the same normalization for the leading and higher twist DAs is used as 
$\psi_{3;{\rm P}}(x)$ defined in Eq.~\eqref{eq:tensor}.

\subsection{Process-Independence}
In this subsection, we shall first compute the decay constants defined by the three different operators 
$\Gamma=(\Gamma_{\rm A}, \Gamma_{\rm P}, \Gamma_{\rm T})$ and show their equivalence, i.e. process-independent decay constant in our LFQM. As the different decay operators are used for the different decay processes, the decay constant's independence of the current operators $\Gamma$ means the independence of the decay process for the decay constant as a physical observable.
The leading-and higher-twist DAs obtained from the matrix elements $\la 0|{\bar q}(z){\Gamma} q(-z)|P\ra$ will be analyzed separately in the next section.

In the LFQM, the decay amplitudes for the operators $\Gamma=(\Gamma_A, \Gamma_{\rm P})$ given by 
Eqs.~(\ref{eq:nonlocal1}) and (\ref{eq:nonlocal2}) can be defined at the level of one-body local current matrix element as
\begin{eqnarray}\label{eq:slf0}
\bra{0} \bar{q}\Gamma q \ket{P} &=& \sqrt{N_c} \int_0^1 {\rm d}x \int \frac{ {\rm d}^2 \mathbf{k}_\bot}{16\pi^3}\  \Phi(x,\mathbf{k}_\perp)  \nonumber\\
   & &  \times \mbox{}
   \sum_{\lambda_1, \lambda_2} {\cal R}_{\lambda_1 \lambda_2}^{00} \left[\frac{\bar{v}_{\lambda_2}(p_2)}{\sqrt{x_2}} \Gamma  \frac{u_{\lambda_1}(p_1)}{\sqrt{x_1}}\right],\quad 
\end{eqnarray}
where $N_c=3$ arises from the color factor implicit in the wave function~\cite{Jaus91,CCH97}. 
Denoting the decay constants $f_{\rm P}$ corresponding to the current operators $(\Gamma_{\rm A}, \Gamma_{\rm P}, \Gamma_{\rm T})$ as $(f_{\rm A}, f_{\rm P}, f_{\rm T})$, we may provide the generic form for the decay constants $f_{\rm A(P)}$ obtained from the two local operators $\Gamma_{\rm A(P)}$ as~\cite{Arifi:2022qnd}
\begin{eqnarray}\label{eq:slf}
f_{\rm A(P)} &=& \sqrt{N_c} \int_0^1 {\rm d}x \int \frac{ {\rm d}^2 \mathbf{k}_\bot}{16\pi^3}\  \Phi(x,\mathbf{k}_\perp)  \nonumber\\
   & &  \times \mbox{} \frac{1}{{\cal P}_{\rm A(P)}}
   \sum_{\lambda_1, \lambda_2} {\cal R}_{\lambda_1 \lambda_2}^{00} \left[\frac{\bar{v}_{\lambda_2}(p_2)}{\sqrt{x_2}}
   \Gamma_{\rm A(P)}  \frac{u_{\lambda_1}(p_1)}{\sqrt{x_1}}\right],
   \nonumber\\
\end{eqnarray}
where we incorporate the Lorentz structures ${\cal P}_{\rm A(P)}= i P^\mu (\mu_M)$ on the RHS of 
Eqs.~(\ref{eq:nonlocal1}) and (\ref{eq:nonlocal2}) into the integral.
This incorporation of Lorentz structures into the integral is to assure the consistent one-body current level of approximation in the computation of the decay constant, which is the crucial aspect of our recently developed LFQM analysis~\cite{C13,C15,Choi:2017uos,Arifi:2022qnd,C21} to obtain the self-consistent, i.e. current-component and boost invariant as well as the process (e.g. $\Gamma_{\rm A(P)}$) independent physical observables by replacing all physical mass $M$ appeared in Eq.~\eqref{eq:slf} with the invariant mass $M_0$.  
So far, most LF calculations of the decay constant, e.g. $f_{\rm A}$ from $\Gamma_{\rm A}$, used $\mu=+$ or $\perp$
since $(P^+, {\bf P}_\perp)$ do not involve any physical mass $M$. On the other hand, the minus component of the
axial-vector current involves $P^-=(M^2 +{\bf P}^2_\perp)/P^+$ and one fails to produce the same result as the one obtained
from the currents with $(\mu=+,\perp)$ if one uses the physical mass in the calculation. The difference between $\mu=-$ and $\mu=(+,\perp)$ was identified as the LF treacherous points such as the instantaneous and 
zero-contributions to the minus current in the solvable covariant model~\cite{C13}. However, in our LFQM
consistent with BT construction, we showed~\cite{Arifi:2022qnd,C21} that the result from the minus component of axial-vector current
gives the same result as the one obtained from the currents with $(\mu=+,\perp)$ if we replace $M$ with $M_0$.
In this work, we shall show that the result obtained from $\Gamma_{\rm P}$ also gives the same result as the
one obtained from $\Gamma_{\rm A}$ as far as we use $M\to M_0$ prescription.  This may be regarded as the effective 
LF zero-mode inclusion at the level of the one-body matrix element calculation
in the LFQM is consistent with BT construction.

For the pseudotensor current ($\Gamma^{\mu\nu}_{\rm T}=\sigma^{\mu\nu}\gamma_5$),
since the decay constant can be computed only in the nonlocal limit (i.e. $z^\mu\neq 0$), 
the calculation incorporating $\psi_{3;{\rm P}}(x)$ is inevitably required in the process of deriving the decay constant from the pseudotensor current.
From the light-like vector $z^\mu$ with $z^+={\bf z}_\perp=0$, 
there are two possible ways to compute the nonlocal matrix element by choosing $\mu\nu=+-$ or $\perp-$. 
\begin{table*}[t]
	\begin{ruledtabular}
		\renewcommand{\arraystretch}{1.6}
		\caption{Various helicity contributions $H_{\lambda_1\lambda_2}$ to the current operators $\mathcal{O}$ for all possible components of the currents, where 
		$\mathcal{A}=(1-x)m_1 + xm_2$, $\mathcal{B}=(1-x)m_1 - xm_2$, $\Delta_1=(m_1^2 +\mathbf{k}_\perp^2)/x$, $\Delta_2=(m_2^2 +\mathbf{k}_\perp^2)/(1-x)$, $\tilde{M}_0^2={(\mathbf{k}_\perp^2 + \mathcal{A}^2)}/{x(1-x)}$, $\mu^0_M=M_0^2/(m_1+m_2)$, and $\rho^0_+=(m_1+m_2)^2/M_0^2$. We should note that the primed symbols in pseudotensor results represent  functions of $x'$ rather than $x$.}
%             }
		\label{tab:coeff2}
		\begin{tabular}{ccccc}
		$\mathcal{O}$	 & Current & $H_{\uparrow\uparrow}+H_{ \downarrow\downarrow}$ & $H_{\uparrow\downarrow}+H_{\downarrow\uparrow}$ 
             & $\mathcal{O}=\sum H_{\lambda_1\lambda_2}$ \\ \hline 
			\multirow{4}{*}{$\mathcal{O}_{\rm A}^\mu$}  & \multirow{2}{*}{$\Gamma^+_{\rm A},\Gamma^\perp_{\rm A}$} & \multirow{2}{*}{$0$} & \multirow{2}{*}{2$\mathcal{A}$} & \multirow{2}{*}{2$\mathcal{A}$}   \\
			& &   \\
		   & \multirow{2}{*}{$\Gamma^-_{\rm A}$} & \multirow{2}{*}{$\dfrac{2(m_1+m_2)\mathbf{k}_\perp^2}{x(1-x)(M_0^2 + \mathbf{P}_\perp^2)}$}   
              & \multirow{2}{*}{$\dfrac{2\mathcal{A}[x(1-x)\mathbf{P}_\perp^2 -\mathbf{k}_\perp^2 + m_1m_2]}{x(1-x)(M_0^2 + \mathbf{P}_\perp^2)}$} 
              & \multirow{2}{*}{$\dfrac{2(m_1\Delta_2 + m_2\Delta_1 + \mathcal{A}\mathbf{P}^2_\perp)}{(M^2_0 + \mathbf{P}^2_\perp)}$}  \\ 
			   & &   \\ \hline 
			  \multirow{2}{*}{$\mathcal{O}_{\rm P}$} &\multirow{2}{*}{$\Gamma_{\rm P}$} &  \multirow{2}{*}{$\dfrac{\mathbf{k}_\perp^2}{x(1-x)\mu^0_M}$}  
                  & \multirow{2}{*}{$\dfrac{\mathcal{A}^2}{x(1-x)\mu^0_M}$ } & \multirow{2}{*}{$\dfrac{\tilde{M}_0^2}{\mu^0_M}$}  \\
			   & &   \\ \hline 
			  \multirow{2}{*}{$\mathcal{O}_{\rm T}^{\mu\nu}$} &\multirow{2}{*}{$\Gamma^{+-}_{\rm T},\Gamma^{\perp -}_{\rm T}$} 
                & \multirow{2}{*}{$\dfrac{(1-2x')\mathbf{k}_\perp^2}{2x'(1-x')(1-\rho^{\prime 0}_+)\mu^{\prime 0}_M} $ } 
                & \multirow{2}{*}{$\dfrac{\mathcal{A'}\mathcal{B'}}{2x'(1-x')(1-\rho^{\prime 0}_+)\mu^{\prime 0}_M} $ } 
                & \multirow{2}{*}{ $\dfrac{-12M'_0k'_z }{\mu^{\prime 0}_M (1-\rho^{\prime 0}_+)} $} \\
                    &  &  \\
		\end{tabular}
		\renewcommand{\arraystretch}{1}
	\end{ruledtabular}
\end{table*}

As for an example, let us choose $\mu\nu = +-$.
We first integrate Eq.~(\ref{eq:tensor}) on both sides using the dummy variable $x^\prime$ (and $\zeta'=2x'-1$) with respect to $z^-$ as
\begin{eqnarray} \label{eq:ten1}
&&\int_{-\infty}^\infty \frac{{\rm d}z^-}{2\pi} {\rm e}^{-i\zeta^\prime P\cdot z}  \bra{0} \bar{q}(z)\Gamma^{+-}_{\rm T} q(-z)\ket{P} \nonumber\\
&&= C P^+ \int_0^1 {\rm d}x\ \int_{-\infty}^\infty \frac{{\rm d}z^-}{2\pi} z^- {\rm e}^{-i(x^\prime-x) P^+z^-}  \psi_{3;{\rm P}}(x),\quad \quad 
\end{eqnarray}
where $C=-\frac{i}{3} f_{\rm T} \left(1 - \rho_+ \right) \mu_M$.
Then, we obtain the RHS of Eq.~\eqref{eq:ten1} via
\bea
&&\int_{-\infty}^\infty \frac{{\rm d}z^-}{2\pi} z^- {\rm e}^{-i(x^\prime-x) P^+z^-}  \psi_{3;{\rm P}}(x)
\nonumber\\
&& = \frac{i}{P^+}\frac{\partial}{\partial x'} \int_{-\infty}^\infty \frac{{\rm d}z^-}{2\pi} {\rm e}^{-i(x^\prime-x) P^+z^-}  \psi_{3;{\rm P}}(x)
\nonumber\\
&& = \frac{i}{P^+}\frac{\partial}{\partial x'}\left[ \delta( (x'-x) P^+) \psi_{3;{\rm P}}(x) \right]
\eea
as follows
\be\label{MRHS}
{\rm RHS\; of\; Eq.~\eqref{eq:ten1} } =\frac{1}{3P^+} f_{\rm T} \left(1 - \rho_+ \right) \mu_M \frac{\partial}{\partial x'}\psi_{3;{\rm P}}(x').
\ee
On the other hand, the left-hand side (LHS) of Eq.~\eqref{eq:tensor} can be rewritten as
\bea\label{MLHS}
&&{\rm LHS\; of\; Eq.~\eqref{eq:ten1} } \nonumber\\
&&=  \sqrt{N_c}\int^1_0 {\rm d}x\int \frac{ {\rm d}^2\mathbf{k}_\perp}{16\pi^3} 
\int_{-\infty}^\infty \frac{{\rm d}z^-}{2\pi} {\rm e}^{-i\zeta^\prime P\cdot z} {\rm e}^{-i(p_2-p_1)\cdot z}
\nonumber\\
&&\times 
\sum_{\lambda_1, \lambda_2}
   \Psi_{\lambda_1 \lambda_2}^{00}(x, {\bf k}_\perp)
  \left[\frac{\bar{v}_{\lambda_2}(p_2)}{\sqrt{x_2}} \Gamma^{+-}_{\rm T}\frac{u_{\lambda_1}(p_1)}{\sqrt{x_1}}\right],
\end{eqnarray}
where ${\rm e}^{-i\zeta^\prime P\cdot z} {\rm e}^{-i(p_2-p_1)\cdot z}={\rm e}^{-i(x'-x)P^+z^-}$ and 
the $z^-$ integration gives the $\delta[(x^\prime - x)P^+]$ and then it is trivially integrated by ${\rm d}x$.
Integrating Eqs.~\eqref{MRHS} and~\eqref{MLHS} over $x'$, we obtain $\psi_{3;P}(x)$ as
\begin{eqnarray}\label{psi3}
\psi_{3;P}(x) &=& \frac{3\sqrt{N_c}}{f_{\rm T}} \int_0^x {\rm d}x^\prime \int \frac{ {\rm d}^2\mathbf{k}_\perp}{16\pi^3}  \Phi(x^\prime,\mathbf{k}_\perp) \nonumber\\
&& \times \frac{1}{{\cal P}_{\rm T}}\sum_{\lambda_1, \lambda_2}
   \mathcal{R'}_{\lambda_1 \lambda_2}^{00} 
  \left[\frac{\bar{v}_{\lambda_2}(p'_2)}{\sqrt{x_2^\prime}} \Gamma^{+-}_{\rm T} \frac{u_{\lambda_1}(p'_1)}{\sqrt{x_1^\prime}}\right], \quad 
  \nonumber\\
\end{eqnarray}
where ${\cal P}_{\rm T}=(1-\rho_+)\mu_M$ and the prime($\prime$) in $({\cal R}, p_i)$ implies that they are functions of $x'$.
By integrating both sides with respect to ${\rm d}x$ and using the normalization of the DA, $\int_0^1 {\rm d}x\ \psi_{3;P}(x)=1$, 
we obtain the decay constant $f_{{\rm T}}$ from the pseudotensor channel as
\begin{eqnarray}\label{fptensor}
f_{\rm T} &=&  3\sqrt{N_c} \int_0^1 {\rm d}x \int_0^x {\rm d}x^\prime \int \frac{ d^2\mathbf{k}_\perp}{16\pi^3}  \Phi(x^\prime,\mathbf{k}_\perp) \nonumber\\
&& \times \frac{1}{{\cal P}_{\rm T}}\sum_{\lambda_1, \lambda_2}
   \mathcal{R'}_{\lambda_1 \lambda_2}^{00} 
  \left[\frac{\bar{v}_{\lambda_2}(p'_2)}{\sqrt{x_2^\prime}} \Gamma^{+-}_{\rm T} \frac{u_{\lambda_1}(p'_1)}{\sqrt{x_1^\prime}}\right].\quad 
\end{eqnarray} 
We should note that the term ${\cal P}_{\rm T}$ including the physical mass $M$ is also incorporated into the integral so that
$M$ is replaced with $M_0$. 
The same results for $\psi_{3;P}(x)$ in Eq.~\eqref{psi3} and $f_{\rm T}$ in Eq.~\eqref{fptensor} can be obtained with $\Gamma^{\perp -}_{\rm T}$. We also note that the main update on the calculation of pseudotensor current compared to Ref.~\cite{Choi:2017uos} is the inclusion of the term $(1-\rho_+)$, 
which leads to the process independence of the decay constant, i.e. $f_{\rm A}=f_{\rm P}=f_{\rm T}$,
as we discuss below. 

Here, we explicitly demonstrate that all three decay constants, $(f_{\rm A}, f_{\rm P}, f_{\rm T})$ as defined by Eqs.~\eqref{eq:slf} and~\eqref{fptensor}, yield identical numerical results.
Using the Dirac helicity spinors~\cite{Lepage:1980fj,Jaus90} and the SO wave function defined in Eq.~\eqref{eq:spinorbit}, 
it is straightforward to compute $(f_{\rm A}, f_{\rm P}, f_{\rm T})$, especially, in terms of
different helicity contributions for different usage of current operators. The final results of $(f_{\rm A}, f_{\rm P}, f_{\rm T})$ in the most general ${\bf P}_\perp\neq 0$ 
frame are summarized as follows
\be\label{eq:operator}
	f_{\rm A(P)} = \sqrt{6}  \int_0^1 {\rm d}x \int \frac{ {\rm d}^2 \mathbf{k}_\bot}{16\pi^3}\  
	\frac{ {\Phi}(x, \mathbf{k}_\bot) }{\sqrt{{\mathcal A}^2 + \mathbf{k}_\bot^2}} ~\mathcal{O}_{\rm A(P)}(x,{\bf k}_\perp),
\ee
and 
\be \label{eq:optensor}
 f_{\rm T} = \sqrt{6} \int_0^1 {\rm d}x \int_0^{x}{\rm d}x^\prime \int \frac{{\rm d}^2\mathbf{k}_\perp }{16\pi^3} \frac{\Phi({x'},\mathbf{k}_\perp) }{\sqrt{{\mathcal{A'}}^2 + \mathbf{k}_\perp^2}} {\mathcal{O}}_{\rm T}(x',{\bf k}_\perp),
 %\quad \quad 
\ee
where ${\cal A'}={\cal A}(x\to x')$. The operators $\mathcal{O}$ given by Eqs.~\eqref{eq:operator} and~\eqref{eq:optensor}
are obtained from the sum of each helicity contribution $H_{\lambda_1\lambda_2}$, i.e.,
\begin{eqnarray}\label{Osum}
   \mathcal{O}=\sum_{\lambda_1,\lambda_2} H_{\lambda_1\lambda_2}.
\end{eqnarray}
The results of each helicity contributions $H_{\lambda_1\lambda_2}$ and their sum ${\cal O}$ defined by Eq.~\eqref{Osum} 
for different current operators 
$\Gamma=(\Gamma_{\rm A},\Gamma_{\rm P}, \Gamma_{\rm T})$
together with different components of the currents for $\Gamma=(\Gamma_{\rm A},\Gamma_{\rm T})$ are summarized in Table~\ref{tab:coeff2}.
We should note that all the physical masses $M$ are replaced with the invariant mass $M_0$ in the final results 
presented in Table~\ref{tab:coeff2}. We confirmed that the three decay constants given by Eqs.~\eqref{eq:operator} and~\eqref{eq:optensor}
are the same as each other, i.e. the pseudoscalar meson decay constant in our LFQM can be obtained in the process-independent 
manner (i.e. $f_{\rm A}=f_{\rm P}=f_{\rm T}$) regardless of the current operators $(\Gamma=\Gamma_{\rm A}, \Gamma_{\rm P}, \Gamma_{\rm T})$ used for the local and nonlocal
matrix elements.

In the Appendix, we also discuss the ``Type II" link~\cite{C13} between the covariant BS model and our
LFQM, which is the alternative method to obtain the self-consistent LFQM results for the decay constants given by Eqs.~\eqref{eq:operator} and~\eqref{eq:optensor}.

\subsection{Lorentz and Rotation Invariance}
In this work, we also compute the decay constant with the nonvanishing $\mathbf{P}_\perp$ frame.
As one can see from Table~\ref{tab:coeff2}, 
the operators ${\cal O}_{\rm P}$, ${\cal O}^{\mu\nu}_{\rm T}$, and ${\cal O}^{+,\perp}_{\rm A}$  
are completely independent of $\mathbf{P}_\perp$. Although the operator ${\cal O}_{\rm A}$ obtained from the minus component of the current $\Gamma^{-}_{\rm A}$
depends on ${\bf P}_\perp$, which is originated from $P^-$ associated with the Lorentz factor $P^\mu$
on the RHS of Eq.~\eqref{eq:nonlocal1}, we confirm that the
decay constant itself is ${\bf P}_\perp$-independent as long as the replacement $M\to M_0$ is made in $P^-$.

In this subsection, we shall explicitly prove not only the ${\bf P}_\perp$-independence but also the rotational
invariance of the decay constant $f_{\rm A(P)}$ given by Eq.~\eqref{eq:operator}.
This can be shown explicitly by converting Eq.~\eqref{eq:operator} into the integral form of the
ordinary three vector ${\vec k}=(k_z, {\bf k}_\perp)$ using Eqs.~\eqref{norm2} and~\eqref{eq:10} 
together with the Jacobi factor given by Eq.~\eqref{Jacob1}, which results in
\be\label{rotinv1}
f_{\rm A(P)} = \sqrt{6}  \int \frac{ {\rm d}^3 {\vec k}}{16\pi^3}\ \sqrt{\frac{M_0}{E_1 E_2}}
	\frac{ {\Phi}({\vec k}) }{{\tilde M_0}} ~\mathcal{O}_{\rm A(P)}({\vec k}),
\ee
where ${\Phi}({\vec k})$ now becomes the wave function mixed with $\phi_{1S}({\vec k})$ and $\phi_{2S}({\vec k})$ 
given by Eq.~\eqref{HO1S2S}. For the pseudoscalar current case, the rotational invariance of the operator 
${\cal O}_{\rm P}={\tilde M}^2_0/\mu^0_{M}$ is evident. 
For the axial-vector current case, the operators ${\cal O}^{(+,\perp)}_{\rm A}=2{\cal A}$ can be converted into
\be\label{OApluskz}
{\cal O}^{(+,\perp)}_{\rm A}({\vec k}) =\frac{2}{M_0} \left[ m_1 E_2 + m_2 E_1 + (m_1 - m_2) k_z \right],
\ee
where the last $k_z$-term vanishes for the $k_z$-integration in Eq.~\eqref{rotinv1} and the rest of terms are
rotationally invariant. Finally, the operator ${\cal O}^-_{\rm A}$ satisfies
\begin{eqnarray}\label{opdiff}
%  \tilde{\mathcal{O}}^{+-}_{\rm PA } =  
  \mathcal{O}^{+}_{\rm A}({\vec k}) -  \mathcal{O}^{-}_{\rm A} ({\vec k)} 
  = \frac{ 4(m_2-m_1) M_0   }{(\mathbf{P}_\perp^2 + M_0^2)}k_z,
\end{eqnarray}
which is an odd function of $k_z$. Equation~\eqref{opdiff} indicates that the decay constant obtained from the minus
current is not only independent of ${\bf P}_\perp$ but also completely equivalent to the one obtained from
the plus (and perpendicular) component of the current. 
It is worth noting that the utilization of a factorized form of the LFWFs, 
such as $\Psi(x)=\Psi_1(x)\Psi_2({\bf k}_\perp)$~\cite{Li:2021cwv}, would result in breaking this rotational invariance.

In the numerical section, we will conduct a quantitative analysis to examine the 
${\bf P}_\perp$-independence of ${\cal O}^-_{\rm A}$ through the ${\bf P}_\perp$-dependence of the helicity
contributions.

\section{Distribution amplitudes}
\label{sec:da_formula}

In this section, we discuss the two-particle DAs up to twist 4 obtained from the three different pseudoscalar meson decay modes.
We summarize in Table~\ref{tab:twist} the twist classification based on the choice of the currents $(\gamma^\mu\gamma_5, i\gamma_5,\sigma^{\mu\nu}\gamma_5)$ 
and all possible components of the currents.

\begin{table}[b]
	\begin{ruledtabular}
		\renewcommand{\arraystretch}{1.5}
		\caption{Twist classification based on the choice of the current and its component.}
		\label{tab:twist}
		\begin{tabular}{cccc}
		Current     & Comp     & Twist  & DAs \\ \hline 
              $\gamma^\mu\gamma_5$	& $+,\perp$	    & 2    & $\phi_{2;{\rm P}}$  \\ 
              & 	$-$	             & 4  & $\phi_{4;{\rm P}}$    \\ 
              $i\gamma_5$ &	 \dots           & 3    & $\phi_{3;{\rm P}}$  \\ 
              $\sigma^{\mu\nu}\gamma_5$	    & $+-,\perp -$	  & 3  & $\psi_{3;{\rm P}}$\\ 
		\end{tabular}
	\end{ruledtabular}
\end{table}

The DAs up to twist-4 accuracy for the pseudoscalar meson with axial-vector current $\Gamma_{\rm A}$ 
are defined in terms of the following matrix element of gauge invariant nonlocal operators as~\cite{Ball:2006wn}
\begin{eqnarray}
 A_{\rm A}^\mu&=& \bra{0} \bar{q}(z)\gamma^\mu\gamma_5 q(-z)\ket{{\rm P}(P)}, \nonumber\\
&= & i f_{\rm A}  \int^1_0 {\rm d}x\ {\rm e}^{i\zeta P\cdot z} \biggl[ P^\mu \left( \phi_{2;{\rm P}}(x) + z^2(\cdots) \right)\nonumber\\
& &\hspace{0.5cm} + \frac{ M^2}{2}\frac{z^\mu}{P \cdot z}   \biggl(\phi_{4;{\rm P}}(x) - \phi_{2;{\rm P}}(x) \biggr)\biggr].
\end{eqnarray}
In order to make a connection between the DAs and the LFWFs of the meson,  we utilize
the equal LF time condition on the light-like vector $z^\mu$ (i.e. $z^2= z^-z^+ -\mathbf{z}_\perp^2=0$) with $z^+ = \mathbf{z}_\perp=0$. 
We then obtain
\begin{eqnarray}\label{eq:DA1}
A^\mu_{\rm A}\biggr|_{z^+ = \mathbf{z}_\perp=0}
&=&  i f_{\rm A} \int^1_0 {\rm d}x\ {\rm e}^{i\zeta P\cdot z} \biggl[ P^\mu\phi_{2;{\rm P}}(x) 
\nonumber\\
& &+ 
\frac{  M^2 z^\mu}{P^+ z^-}  \biggl(\phi_{4;{\rm P}}(x)-\phi_{2;{\rm P}}(x) \biggr) \biggr].\quad \quad 
\end{eqnarray}
To isolate the twist-2 DA, $\phi_{2;{\rm P}}(x)$, one may take either the plus
or transverse component of the current and obtain
\be
A^{(+,\perp)}_{\rm A}
=  i f_{\rm A} P^{(+,\perp)} \int^1_0 {\rm d}x\ {\rm e}^{i\zeta P\cdot z}  \phi_{2;{\rm P}}(x).\quad \quad 
\ee
This explains why the two decay constants, $f^{(+)}_{\rm A}$ and $f^{(\perp)}_{\rm A}$, have the same operator
${\cal O}^{(+)}_{\rm P} = {\cal O}^{(\perp)}_{\rm P}$.
On the other hand, the twist-4 DA $\phi_{4;{\rm P}}(x)$ can be obtained from the minus component of the current in the ${\bf P}_\perp=0$ frame as
\be
A^-_{\rm A} = i f_{\rm A} P^- \int^1_0 {\rm d}x\ {\rm e}^{i\zeta P\cdot z}  \phi_{4;{\rm P}}(x).\quad
\ee
Here it is shown that the higher twist DAs are associated to the bad current while the leading twist DAs correspond to the good current.

%%% Table 1
\begin{table*}[t]
	\begin{ruledtabular}
		\renewcommand{\arraystretch}{1.3}
		\caption{The constituent quark masses $m$ [GeV] and variational parameters $\beta$ [GeV] for $\theta=12^\circ$ adapted from our previous work~\cite{Arifi22}.  }
		\label{tab:parameter}
		\begin{tabular}{ccccccccccc}
			 $m_q$ & $m_s$  & $m_c$ & $m_b$ & 
			$\beta_{q\bar{c}}$ & $\beta_{s\bar{c}}$  & $\beta_{q\bar{b}}$ & $\beta_{s\bar{b}}$ & $\beta_{c\bar{c}}$ & $\beta_{c\bar{b}}$ & 
			$\beta_{b\bar{b}}$\\ \hline
			 0.22 & 0.45 & 1.68 & 5.10 & 0.424 & 0.455 & 0.495 & 0.538 & 0.592 & 0.767 & 1.167\\
		\end{tabular}
		\renewcommand{\arraystretch}{1}
	\end{ruledtabular}
\end{table*}
%%%

For the twist-3 case, there are two different DAs that are related to pseudoscalar ($\Gamma_{\rm P}$) and pseudotensor ($\Gamma_{\rm T}$) currents. 
For pseudoscalar current, the twist-3 DA $\phi_{3;{\rm P}}(x)$ is uniquely determined by~\cite{Ball:2006wn}
\begin{eqnarray}
A_{\rm P}\biggr|_{z^+ = \mathbf{z}_\perp=0} &=& \bra{0} \bar{q}(z)i\gamma_5 q(-z)\ket{{\rm P}(P)}, \nonumber\\
&=& f_{\rm P} \mu_M \int_0^1 {\rm d}x\ {\rm e}^{i\zeta P\cdot z}  \phi_{3;{\rm P}}(x),
\end{eqnarray}
without choosing a particular component of current.
For pseudotensor current, the DA is computed as~\cite{Ball:2006wn} 
\begin{eqnarray}
A^{\mu\nu}_{\rm T}\biggr|_{z^+ = \mathbf{z}_\perp=0} &=& \bra{0} \bar{q}(z)\sigma^{\mu\nu}\gamma_5 q(-z)\ket{{\rm P}(P)}, \nonumber\\
    &=& -\frac{i}{3} f_{\rm T} \left(1 - \rho_+ \right) \mu_M (P^\mu z^\nu -P^\nu z^\mu) \nonumber\\
    & & \times \int_0^1 {\rm d}x\ {\rm e}^{i\zeta P\cdot z}  \psi_{3;{\rm P}}(x).\quad \quad 
\end{eqnarray}
In this case, the nonvanishing components are $\mu\nu=+-$ and $\perp-$ and 
they give the same $\psi_{3;{\rm P}}(x)$ as we have shown for the computation of the decay constant $f_{\rm T}.$

In our notation, all DAs $\phi_{n;{\rm P}}(x)(n=2,3,4)$ and $\psi_{3;{\rm P}}(x)$ are normalized to unity as
\begin{eqnarray}\label{DAnorm}
    \int_0^1 {\rm d}x\ \left\{\phi_{n; {\rm P}} (x), \psi_{3;{\rm P}}(x)\right\} = 1.
\end{eqnarray}

From Eqs.~\eqref{eq:operator} and~\eqref{eq:optensor} together with Eq.~\eqref{DAnorm},
we obtain $\phi_{n;{\rm P}}(x)(n=2,3,4)$ from the axial-vector ($n=2,4$) and pseudoscalar $(n=3)$ channels as
\begin{eqnarray}
    \phi_{n;\rm P}(x) = \frac{\sqrt{6} }{f_{\rm P}}  \int \frac{{\rm d}^2 \mathbf{k}_\bot}{16\pi^3}\ \frac{\Phi(x,\mathbf{k}_\perp) }{\sqrt{\mathcal{A}^2 + \mathbf{k}_\perp^2} }\   
                                                    \mathcal{O}_{\rm A(P)}.
\end{eqnarray}
Here, we have $\mathcal{O}^{+}_{\rm A}=\mathcal{O}^{\perp}_{\rm A}$ corresponding to $n=2$, 
$\mathcal{O}^{-}_{\rm A}$ corresponding to $n=4$, and $\mathcal{O}_{\rm P}$ corresponding to $n=3$.
For $\psi_{3;{\rm P}}(x)$ from the pseudotensor channel, we obtain
\begin{eqnarray}
    \psi_{3;\rm P}(x) = \frac{\sqrt{6} }{f_{\rm P}} \int^x_0{\rm d}x' \int \frac{{\rm d}^2 \mathbf{k}_\bot}{16\pi^3}\ \frac{\Phi(x',\mathbf{k}_\perp) }{\sqrt{\mathcal{A}^{\prime 2} + \mathbf{k}_\perp^2} }\   \mathcal{O}_{\rm T},
\end{eqnarray}
where $\mathcal{O}_{\rm T}=\mathcal{O}^{+-}_{\rm T}=\mathcal{O}^{\perp -}_{\rm T}$.

\section{Numerical results and discussion}
\label{sec:result}

The model parameters $m$ and $\beta$ used in the present work are summarized in Table~\ref{tab:parameter}, which were determined from the spectroscopic study in our previous work~\cite{Arifi22}. 

\subsection{Light-front wave function}

\begin{figure}[b] 
	\centering
	\includegraphics[width=0.9\columnwidth]{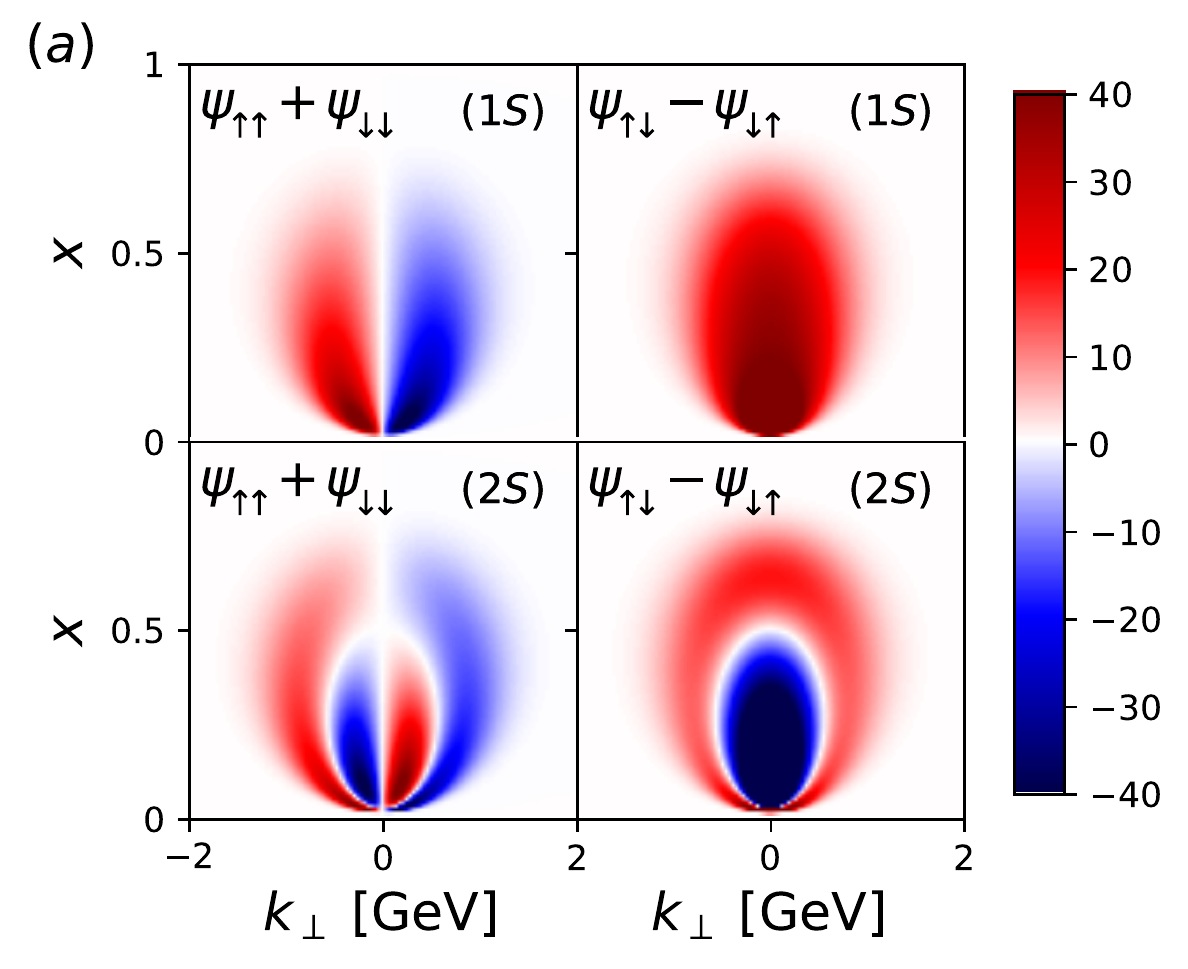}
        \includegraphics[width=0.9\columnwidth]{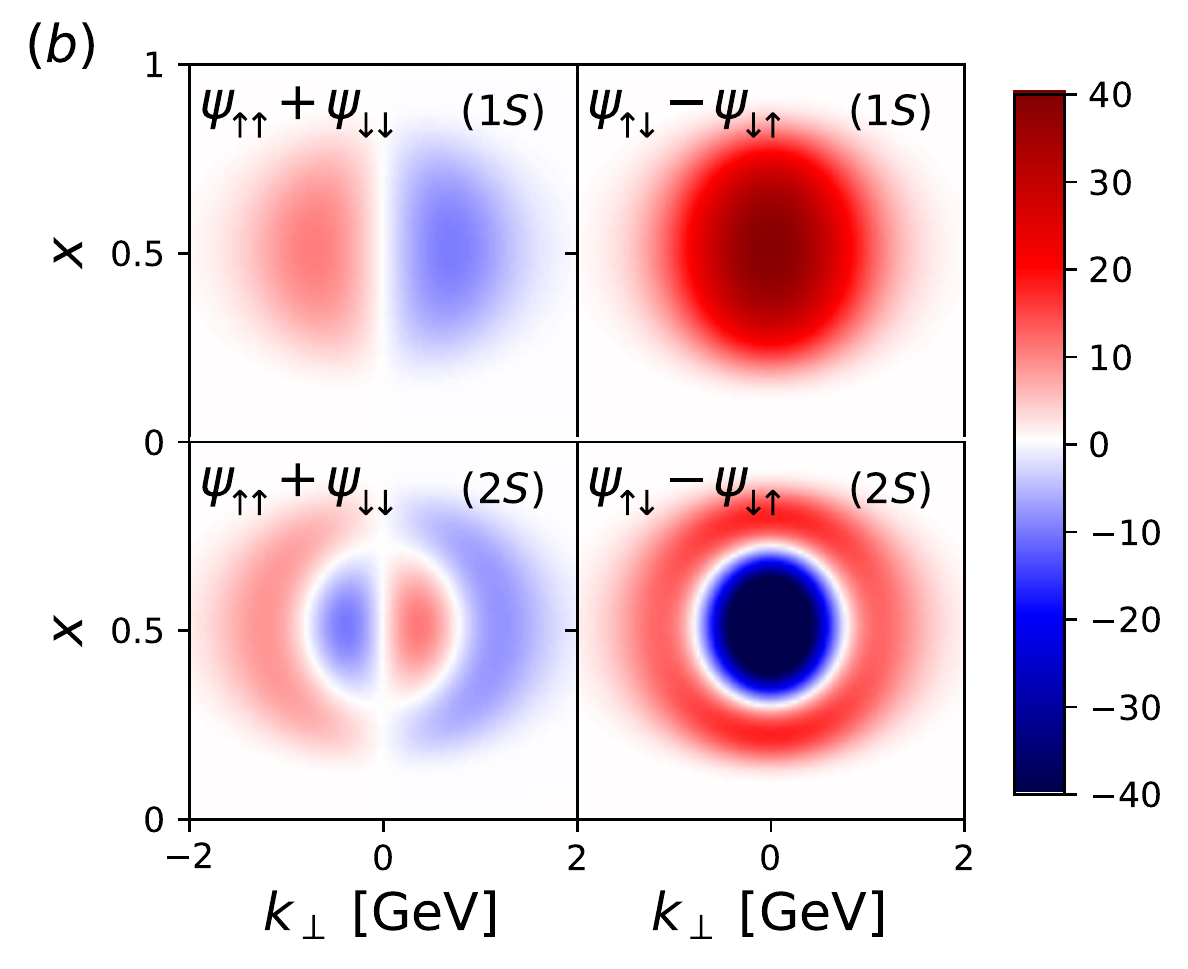}
	\caption{ \label{fig:wavefunction1} Two-dimensional plot of LFWF of (a) $D$ and (b) $\eta_c$ mesons for each helicity configuration. Note that the longitudinal momentum $x$ is carried by the light quark. }
\end{figure}

We first discuss the LFWFs $\Psi^{00}_{\lambda_1\lambda_2}(x, {\bf k}_\perp)$ defined in Eq.~\eqref{eq:5} for the 
ground ($1S$) state and the radially excited ($2S$) state heavy pseudoscalar mesons.
Figure~\ref{fig:wavefunction1}(a) shows the two-dimensional (2D) plots of $D(1S)$ and $D(2S)$ mesons as a function of $(x, k_\perp)$, 
respectively, as an example of the unequal-mass case while Figure~\ref{fig:wavefunction1}(b) shows the 2D plots of $\eta_c(1S)$ and $\eta_c(2S)$ heavy quarkonia, 
respectively, as an example of the equal-mass case.
In Fig.~\ref{fig:wavefunction1}, the LFWFs $\Psi^{00}_{\lambda_1\lambda_2}$ are
presented in term of the helicity configuration $\lambda_1\lambda_2$ where we denote $\lambda=+1/2$ and $-1/2$ as $\uparrow$ and $\downarrow$, respectively.
Note that the longitudinal momentum fraction $x$ is carried by the lighter quark with mass $m_1$. 
The LFWFs can be compared with those obtained in Ref.~\cite{Li:2021cwv}.

\begin{figure*}[t]
	\centering
    \includegraphics[width=1\columnwidth]{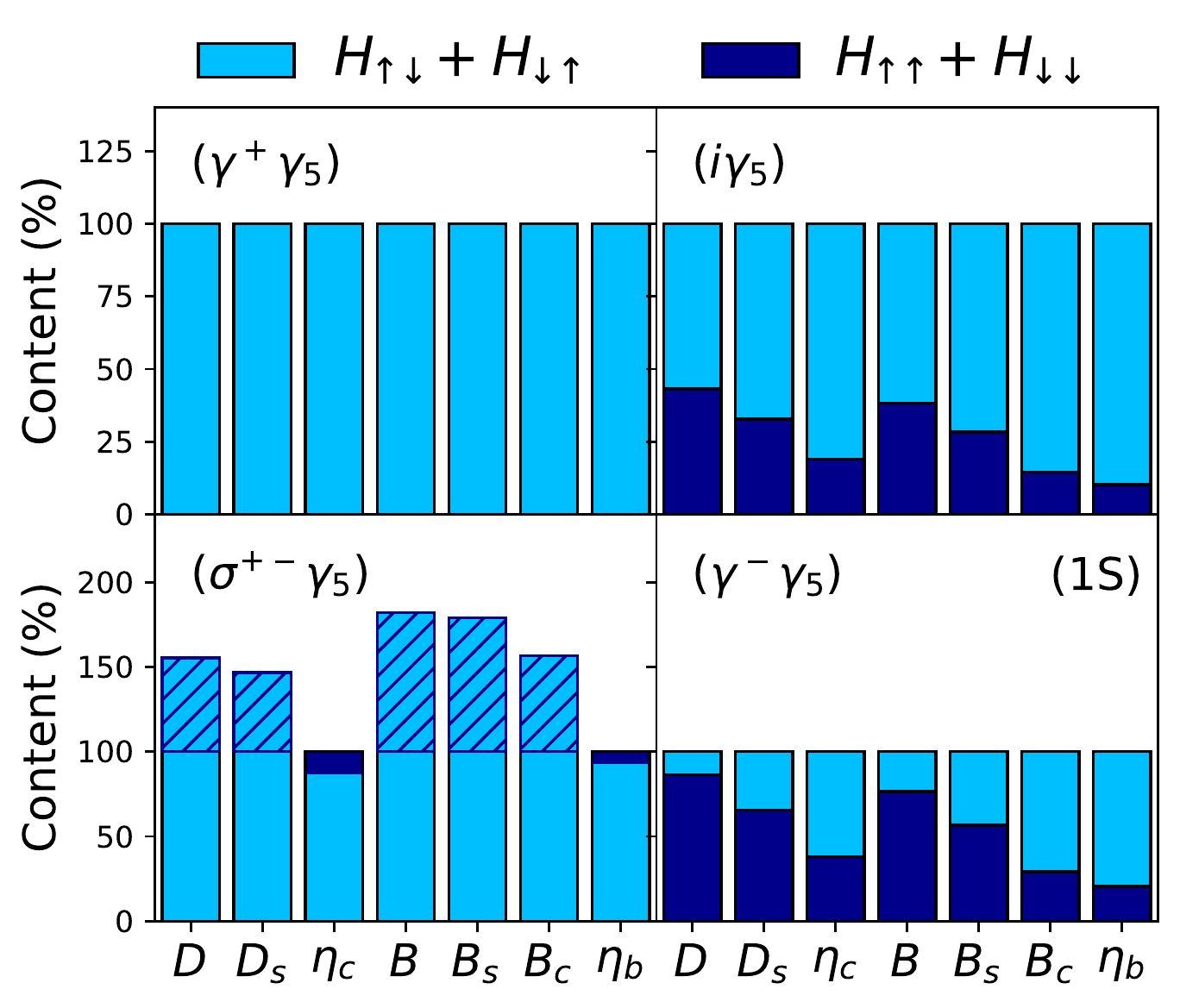}
    \includegraphics[width=1\columnwidth]{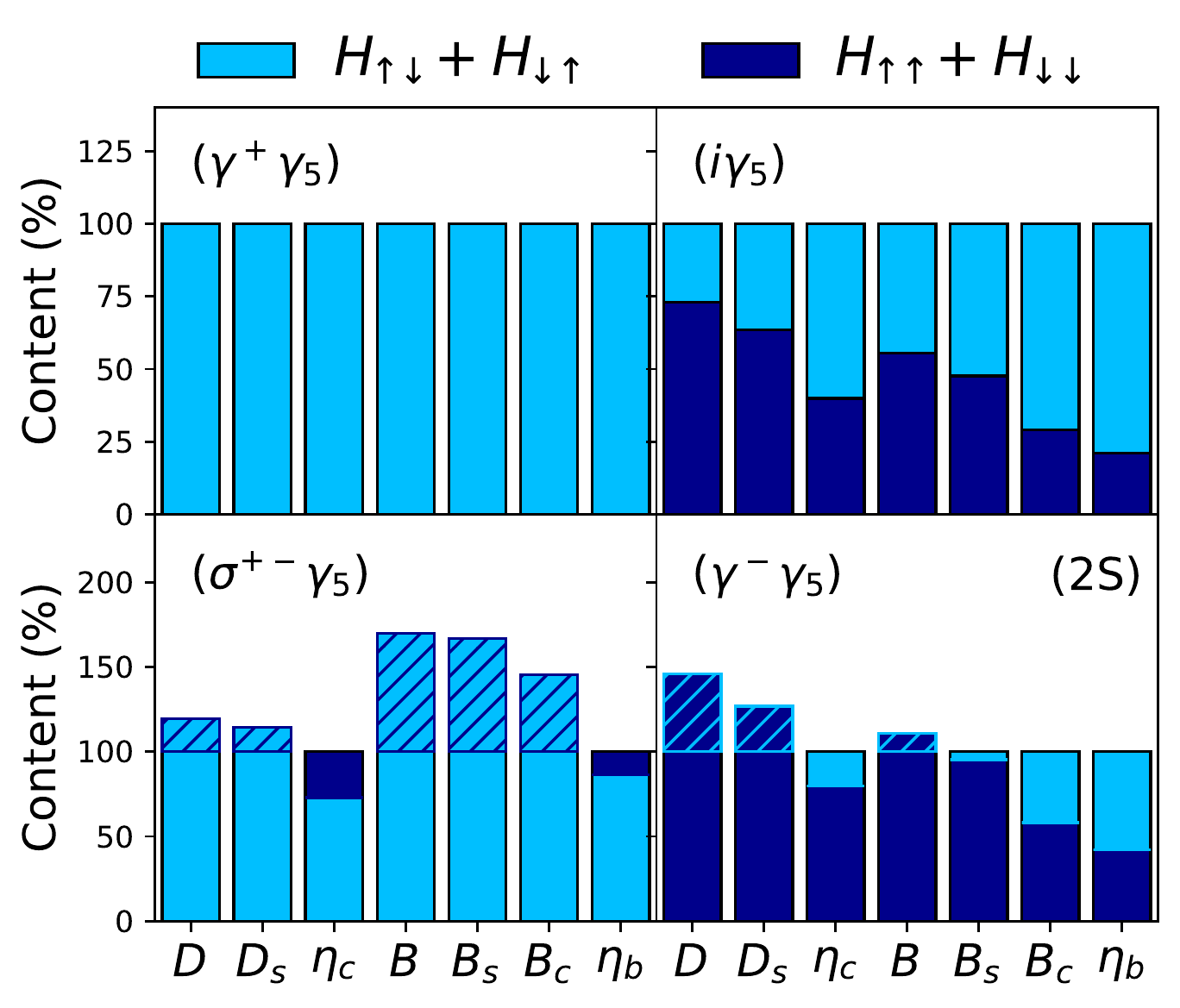}
	\caption{ Helicity contributions to decay constant for 1S and 2S heavy pseudoscalar mesons. The pattern histogram represents subtracted contribution. Since the contribution from $H_{\uparrow\downarrow}=H_{\downarrow\uparrow}$, we sum them up for simplicity. It also applies to $H_{\uparrow\uparrow}=H_{\downarrow\downarrow}$. Here we set $\mathbf{P}_\perp=0$ for the case of $\gamma^-\gamma_5$. For the $\sigma^{\mu\nu}\gamma_5$ case, the helicity contribution depends on the choice of assigning $x$ to light or heavy quark, see Sec.~\ref{sec:da}, for the detail.\label{fig:helicity}}
\end{figure*}

\begin{figure*}[t]
	\centering
    \includegraphics[width=2\columnwidth]{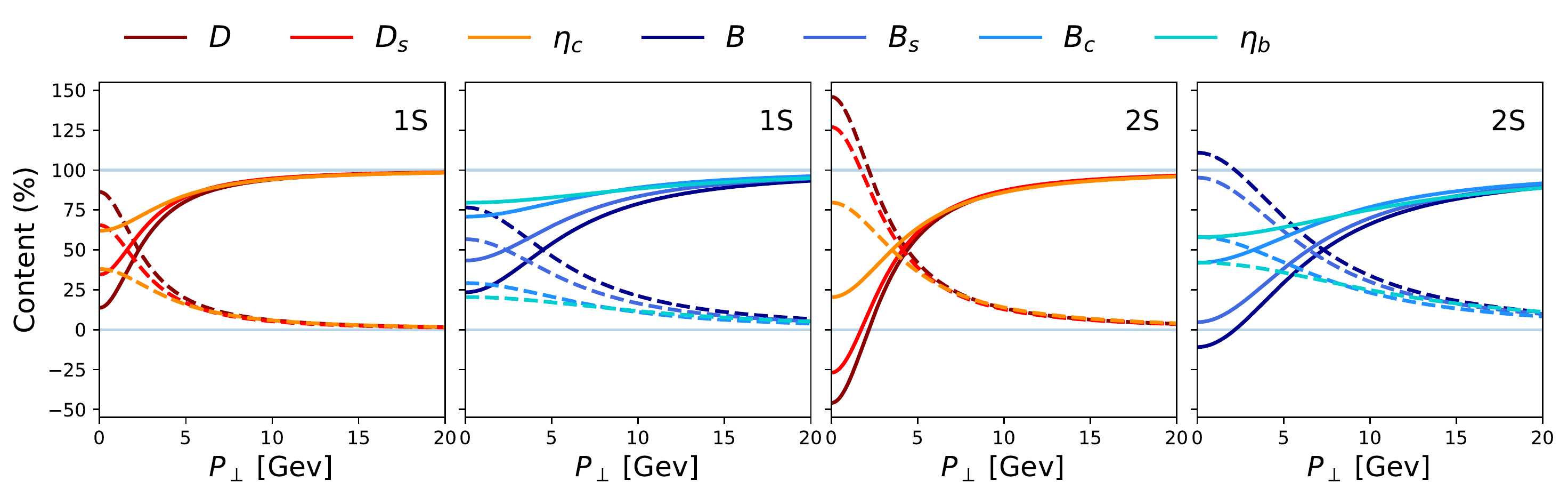}
	\caption{ The $P_\perp$ dependence of the helicity contribution of $\gamma^-\gamma_5$ for 1S and 2S pseudoscalar heavy mesons with $|\mathbf{P}_\perp|=P_\perp$. The solid line represents the contributions from $H_{\uparrow\downarrow}+H_{\downarrow\uparrow}$ while the dashed line represents $H_{\uparrow\uparrow}+H_{\downarrow\downarrow}$. \label{fig:pt}}
\end{figure*}

There are several salient features related to the LFWFs in Fig.~\ref{fig:wavefunction1}. 
(i) The center of the LFWF ($k_\perp \to 0$ and $k_z\to 0$), which is associated with its extremum point, is located at 
\begin{eqnarray}
    x = \frac{m_1}{m_1 + m_2},
\end{eqnarray}
which is obtained by solving Eq.~(\ref{eq:kz}).
For the equal-mass case, $\Psi(k_\perp \to 0, k_z\to 0)$ is located at $x=1/2$ as can be seen in Fig.~\ref{fig:wavefunction1}. 
But, for the unequal-mass case (i.e. $m_1=m_{u(d)}, m_2=m_c$) in Fig.~\ref{fig:wavefunction1}, 
the center moves to the value of $x<1/2$ and therefore, the LFWF is somewhat distorted on $(x, k_\perp)$ plane. 
(ii) The LFWF correctly represents the pseudoscalar meson as $\Psi_{\uparrow\downarrow}^{00}(x,\mathbf{k}_\perp) = -\Psi_{\downarrow\uparrow}^{00}(x,\mathbf{k}_\perp).$
In addition to the ordinary helicity $(\uparrow\downarrow,\downarrow\uparrow)$, there is also a nonvanishing contribution from the higher helicities $(\uparrow\uparrow, \downarrow\downarrow)$ that couple to the quark orbital angular momentum as the sign is different in the positive and negative domain of $k_\perp$. 
This configuration is possible in the relativistic dynamics.
However, such contribution is suppressed as the quark mass increases and vanishes in the heavy-quark limit ($m\to\infty$).
Therefore, at the heavy-quark or nonrelativistic limit, the LFWF will take only the contribution from the ordinary helicity without involving the orbital angular momentum.
(iii) It is also shown that the $2S$ state has a nodal structure represented as a white circle/oval where the center of LFWF has a dip represented as a blue region for the case of ordinary helicity. One may notice that the shape of the LFWFs is largely reflected in the DAs. 

\subsection{Decay constant}

\begin{figure}[b]
	\centering 
    \includegraphics[width=0.9\columnwidth]{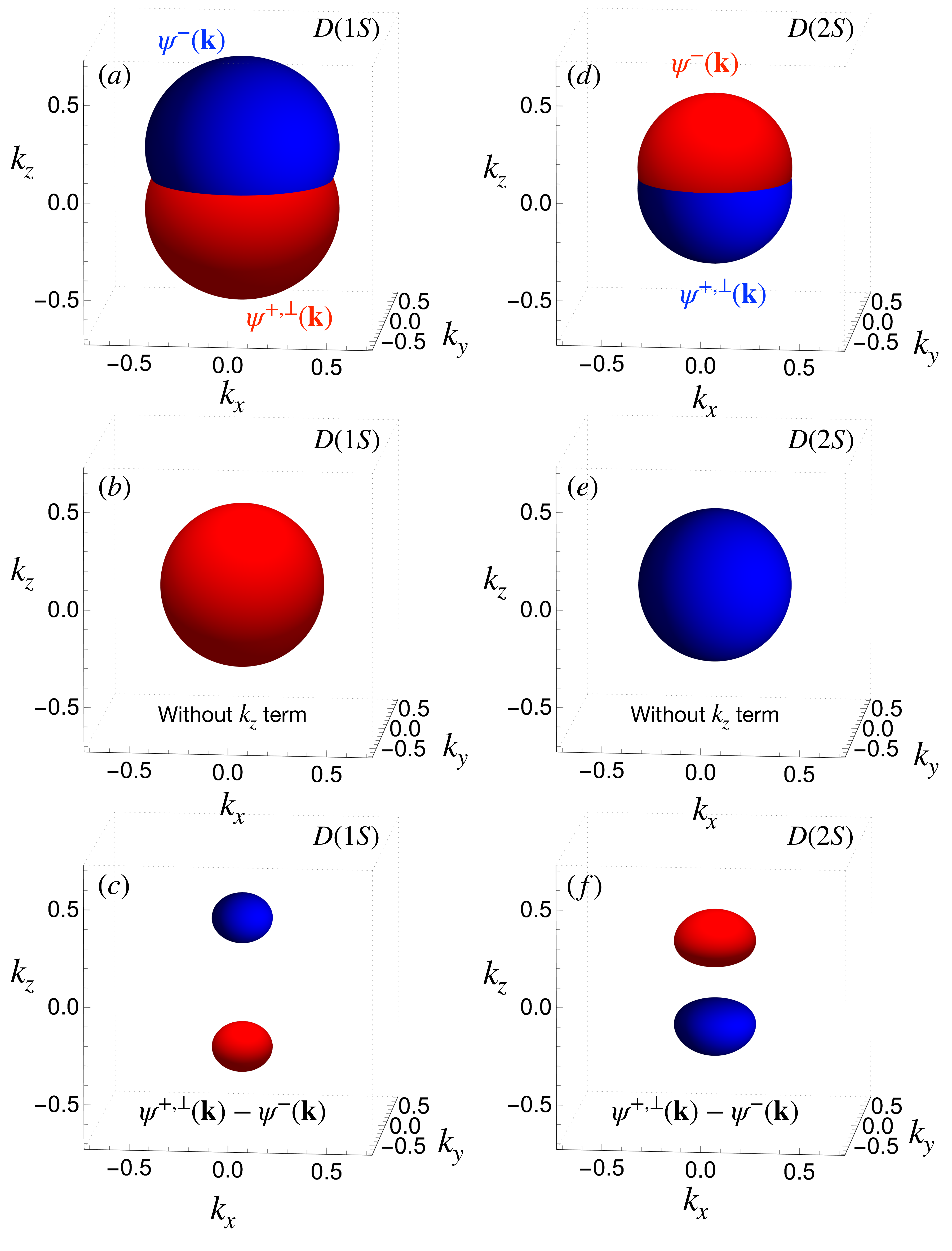}
	\caption{ \label{fig:3d_constant} The 3D plots of wave functions for $D(1S)$ mesons defined by $f_{A}=\int {\rm d}^3\vec{k}\ \psi^\mu(\vec{k})$: (a) $\psi^\mu(\vec{k})$ with various current component, (b) $\psi^\mu(\vec{k})$ without the $k_z$ term, and (c) $\tilde{\psi} (\vec{k})=\psi^{+,\perp}(\vec{k})- \psi^-(\vec{k})$. Displayed in (d), (e), and (f) is for those of $D(2S)$ mesons.}
\end{figure}

%%%
\begin{table}[b]
	\begin{ruledtabular}
		\renewcommand{\arraystretch}{1.3}
		\caption{Decay constants of heavy pseudoscalar mesons predicted in the LFQM~\cite{Arifi22}. The results are given in MeV. }
		\label{tab:constant}
		\begin{tabular}{ccc|ccc} 
             State   & $f_{theo}$ & $f_{exp}$ & State   & $f_{theo}$ & $f_{exp}$ \\ \hline
            $D(1S)$ & 208 & 206.7(8.9) & $D(2S)$ & 110 & \dots \\
            $D_s(1S)$ & 246 & 257.5(6.1) & $D_s(2S)$ & 133 & \dots  \\
             $\eta_c(1S)$ & 348 & 335(75) & $\eta_c(2S)$ & 214 & \dots \\
            $B(1S)$ & 190 & 188(25) & $B(2S)$ & 126 & \dots \\
            $B_s(1S)$ & 228 & \dots & $B_s(2S)$ & 150 & \dots  \\
             $B_c(1S)$ & 394 & \dots & $B_c(2S)$ & 268 & \dots \\
             $\eta_b(1S)$ & 628 & \dots & $\eta_b(2S)$ & 443 & \dots\\ 
      \end{tabular}
		\renewcommand{\arraystretch}{1} 
	\end{ruledtabular}
\end{table}
%%%

First of all, the numerical values of decay constants for $1S$ and $2S$ state heavy pseudoscalar mesons obtained from the axial-vector current are presented 
in our previous work~\cite{Arifi22}.
In this work, we confirm that the decay constants obtained from the pseudoscalar and pseudotensor channels also produce the same results as those obtained from the
axial-vector channel
regardless of the currents as well as all possible current components. Namely, we obtain the process-independent pseudoscalar meson decay constants in the LFQM.
For the sake of completeness, we display again the results of $1S$ and $2S$ state heavy pseudoscalar mesons for the case of mixing angle $\theta=12^\circ$
in Table~\ref{tab:constant}.

In addition, we examine in Fig.~\ref{fig:helicity} the contributions of helicity to the decay constants for $1S$ and $2S$ state heavy pseudoscalar mesons, 
as they are contingent upon the current component, 
as indicated in Table~\ref{tab:coeff2}. 
Notably, as observed in Fig.~\ref{fig:helicity}, the helicity contributions exhibit variation across different currents and current components. 
Despite these variations, however, the resulting decay constant remains unaltered.

For the $\gamma^+\gamma_5$, the contribution denoted by $H_{\uparrow\downarrow} + H_{\downarrow\uparrow}$ is entirely from the ordinary helicity wave function $\Psi_{\uparrow\downarrow-\downarrow\uparrow}^{00}(x,\mathbf{k}_\perp)$ without involving the orbital angular momentum.
This is one of the reasons why we call plus current $(\mu=+)$ as the good current where the dynamics becomes much simpler and it is also related to the leading twist DAs as explained in Sec.~\ref{sec:da_formula}. For the $\gamma^\perp\gamma_5$, the contribution is still entirely from the ordinary helicity. However, when we use $\gamma^-\gamma_5$ or the bad current, the higher helicity contributions denoted by $H_{\uparrow\uparrow} + H_{\downarrow\downarrow}$ arise and the dynamics becomes more complicated. 
It is clearly shown that the higher helicity contribution is suppressed when the constituent quark mass becomes heavier. 
A similar behavior is also observed for the $i\gamma_5$ case. 
When we use $\sigma^{+-}\gamma_5$ (or $\sigma^{\perp-}\gamma_5$), the ordinary helicity contribution appears more than expected as shown in the bottom left panel of Fig.~\ref{fig:helicity} for some cases.
However, the higher helicity contribution compensates for it and keeps the decay constant remains the same.
It is also worth mentioning that the behaviors of helicity contribution for the ground state and the radially excited state are similar. 
The difference is that the higher helicity contribution is more pronounced for the radially excited state. 

In Fig.~\ref{fig:pt}, we show the ${\bf P}_\perp$-independence of the decay constants for the $1S$ and $2S$ state heavy pseudoscalar mesons. 
While each helicity contribution shows the $\mathbf{P}_\perp$ dependence when one uses the minus component of the axial-vector current, the sum of all helicity
contributions is completely independent of ${\bf P}_\perp$ as it should be.
It is also evident from Fig.~\ref{fig:pt} that the higher (ordinary)  helicity contributions dominate at low (high) $\mathbf{P}_\perp$ region consistent with the previous
observation for the equal quark mass case~\cite{Arifi:2022qnd}.
We can also see that higher helicity is more enhanced for the $2S$ state, similar to that in Fig.~\ref{fig:helicity}. 

\begin{figure*}[t]
	\centering 
    \includegraphics[width=1.9\columnwidth]{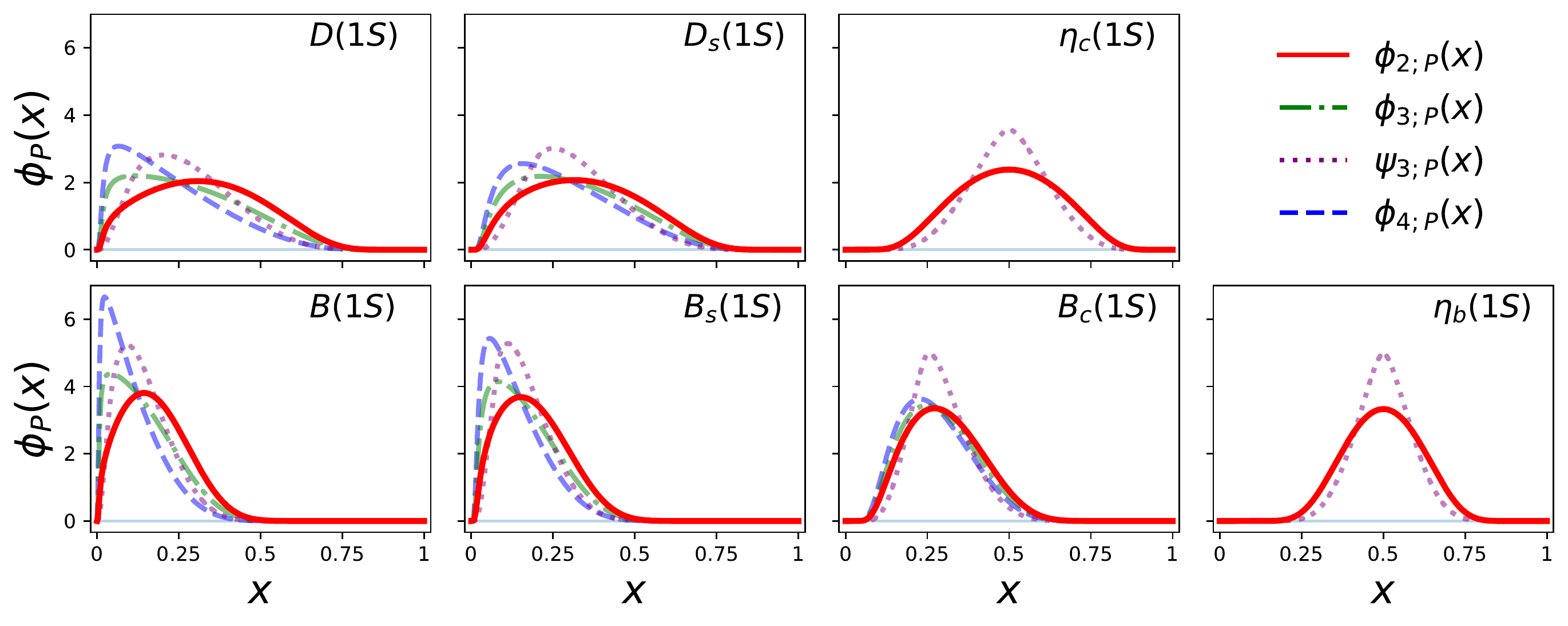}
    \includegraphics[width=1.9\columnwidth]{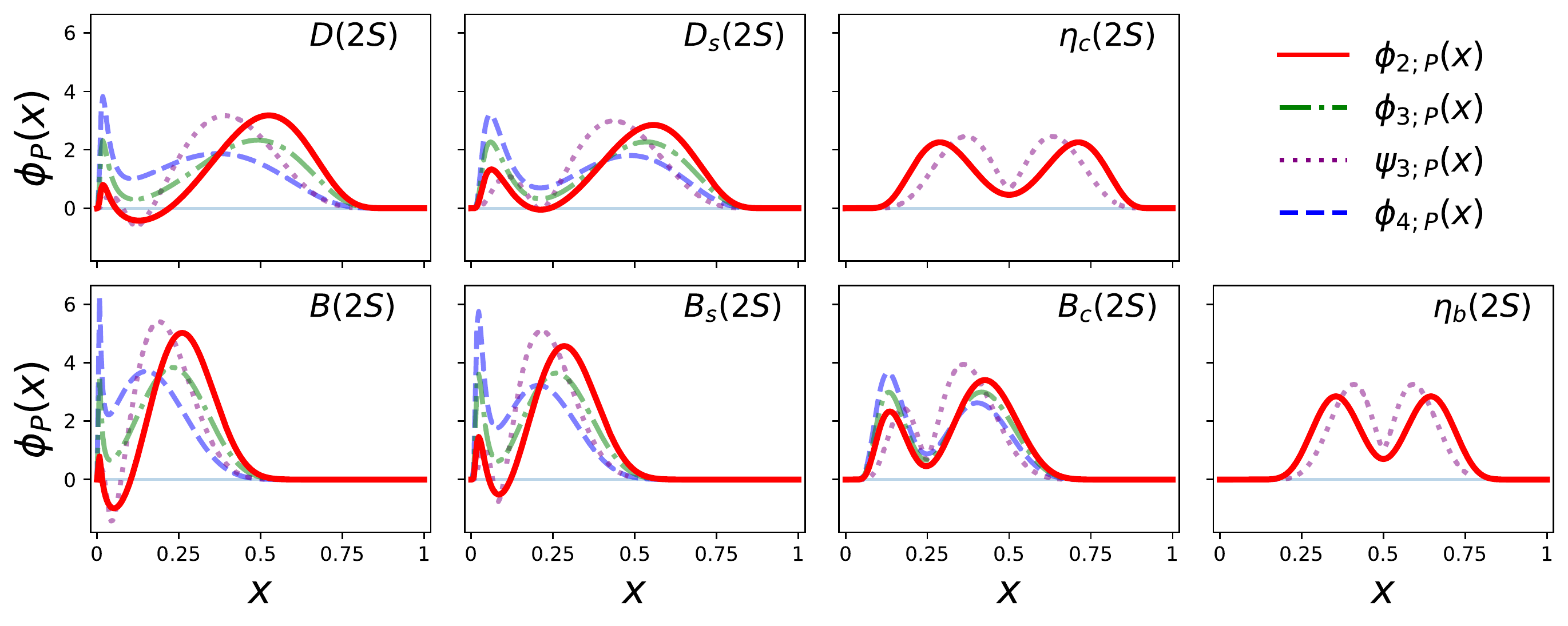}
	\caption{ \label{fig:da} Two-particle DAs with various twists for 1S and 2S heavy pseudoscalar mesons with various quark flavor contents where the longitudinal momentum $x$ is carried by the light quark.}
\end{figure*}

Finally, we also examine the rotational invariance of the decay constant by investigating the wave function $\psi^\mu(\vec{k})$ defined in Eq.~\eqref{rotinv1} 
with $f_{A}=\int {\rm d}^3\vec{k}\ \psi^\mu(\vec{k})$ for the axial-vector $(\Gamma^\mu_{\rm A}=\gamma^\mu\gamma_5)$ current.
For the equal-mass case, the $\psi^\mu(\vec{k})$ has a spherical shape since the obtained operator $\mathcal{O}_A^\mu = 2m$ regardless of the current component $\mu$. 
For the unequal-mass case such as $D$ meson, the wave functions $\psi_D^{+,\perp}(\vec{k})$ and $\psi_D^{-}(\vec{k})$ are slightly deformed and shifted to the negative and positive $k_z$ domains\footnote{The shifting to either positive or negative domain depends on the choice of $m_1$ and $m_2$ since $\mathcal{O}_A^-(\vec{k}) \propto (m_1-m_2)k_z$ in Eq.~\eqref{OApluskz}. }, respectively, as depicted in the upper panels of Fig.~\ref{fig:3d_constant}.
It is also shown that the wave functions $\psi^{\mu}(\vec{k})$ are more separated in the $k_z$ direction for the $D(1S)$ state compared to those of the $D(2S)$ state. 
The shifting in $k_z$ direction can be understood by the appearance of the odd function of $k_z$ for ${\cal O}_A^{+,\perp}(\vec{k})$ in Eq.~\eqref{OApluskz} 
although such an odd $k_z$ term does not actually contribute to the decay constant.
The wave functions $\psi^{+,\perp}(\vec{k})$ and $\psi^{-}(\vec{k})$ become a sphere centered at the origin and coincide with each other if the $k_z$ term is removed as shown in the middle panels of Fig.~\ref{fig:3d_constant}. 
Moreover, the difference defined as $\tilde{\psi}_D(\vec{k})=\psi^+_D(\vec{k})-\psi^-_D(\vec{k})$ is displayed in lower panels of Fig.~\ref{fig:3d_constant} showing that the integration over $k_z$ will give a vanishing result.
Therefore, it is evident that the decay constant with $\mu=+,\perp,-$ is the same.
As for the pseudoscalar current, the wave functions are spherical as also implied from its operator ${\cal O}_P(\vec{k})$.

\subsection{Distribution amplitude}
\label{sec:da}

\begin{figure*}[t]
	\centering 
    \includegraphics[width=1.9\columnwidth]{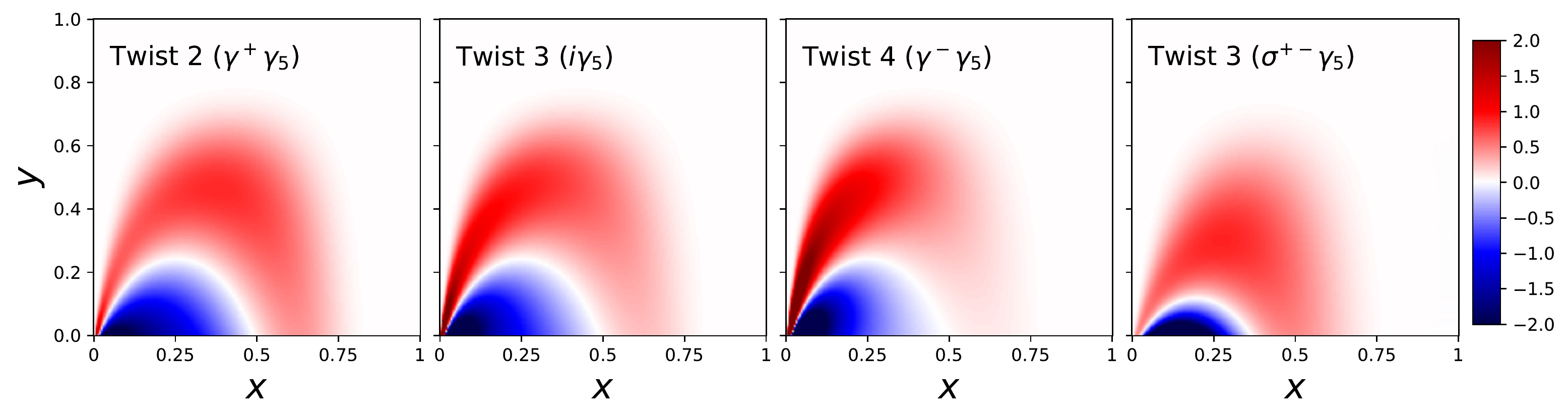}
	\caption{ \label{fig:da_2d} Two-dimensional plot of the the DAs of $D(2S)$ for various twist. Here we define ${\bf k}_\perp^2 = y/(1-y)$ to make a rectangular domain.  The longitudinal momentum $x$ is carried by the light quark.}
\end{figure*}

\begin{figure*}[t]
	\centering 
    \includegraphics[width=1.9\columnwidth]{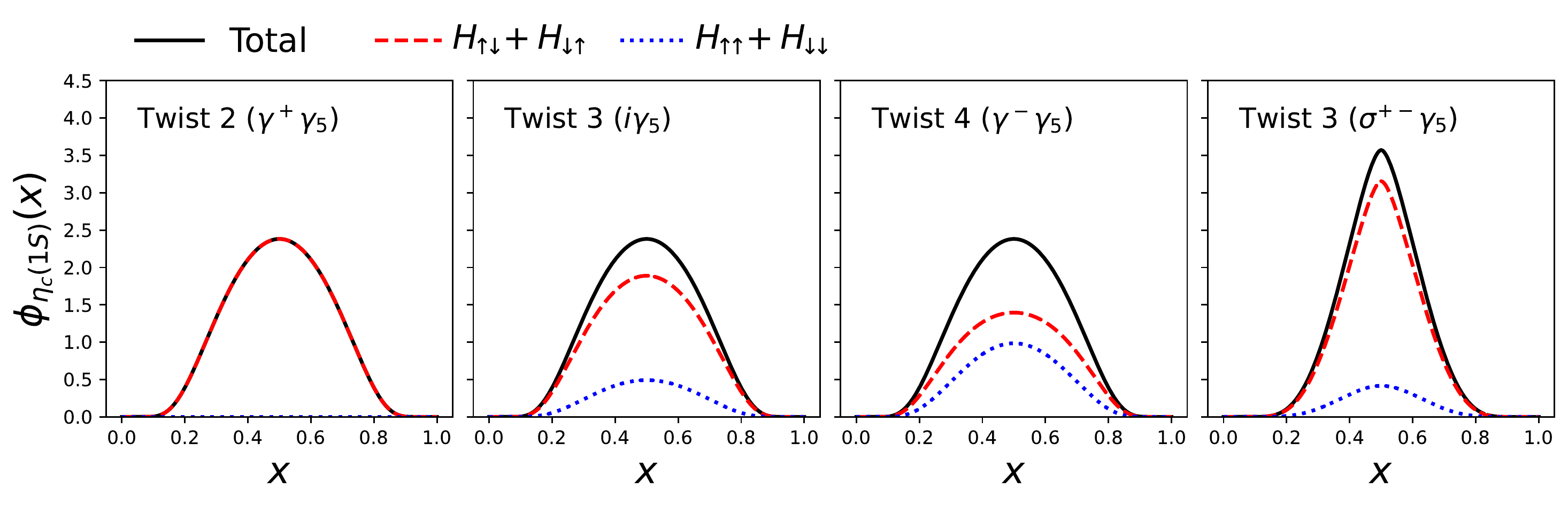}
	\caption{ \label{fig:da_hel} Helicity contribution to DAs with various twist. Although the total DAs of twist 2, twist 3, and twist 4 are the same, they consist of different helicity contributions. We note that the total DA of twist 3 with $\sigma^{+-}\gamma_5$ is different from the others. }
\end{figure*}

\begin{figure}[t]
	\centering 
    \includegraphics[width=0.9\columnwidth]{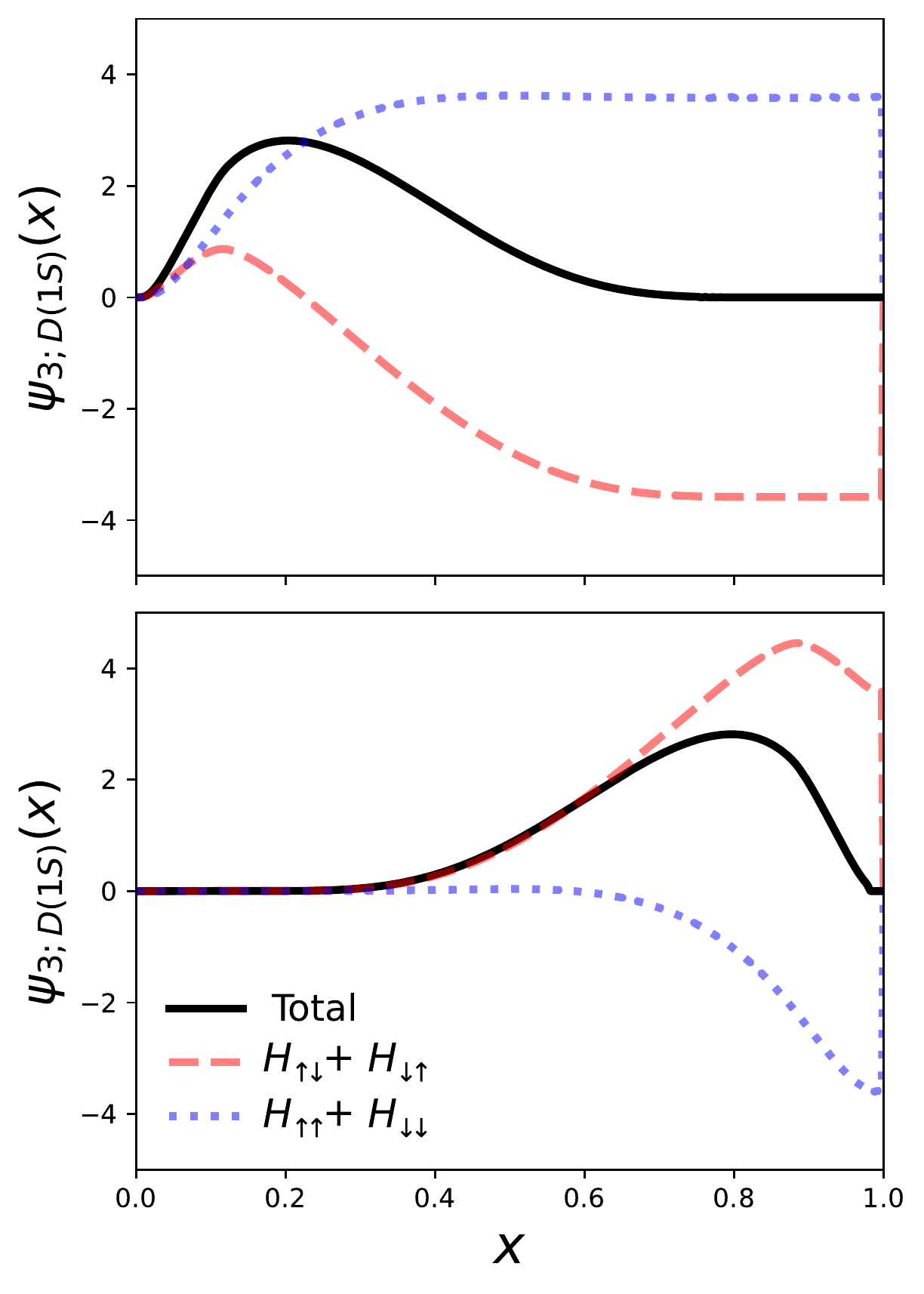}
	\caption{ \label{fig:da3t_hel} The difference in the helicity contribution depending on the choice of assigning the LF longitudinal momentum fraction $x$ to the light or heavy quark. 
 The dashed and dotted lines represent the ordinary and higher helicity, respectively. }
\end{figure}

\begin{figure}[t]
	\centering 
    \includegraphics[width=0.9\columnwidth]{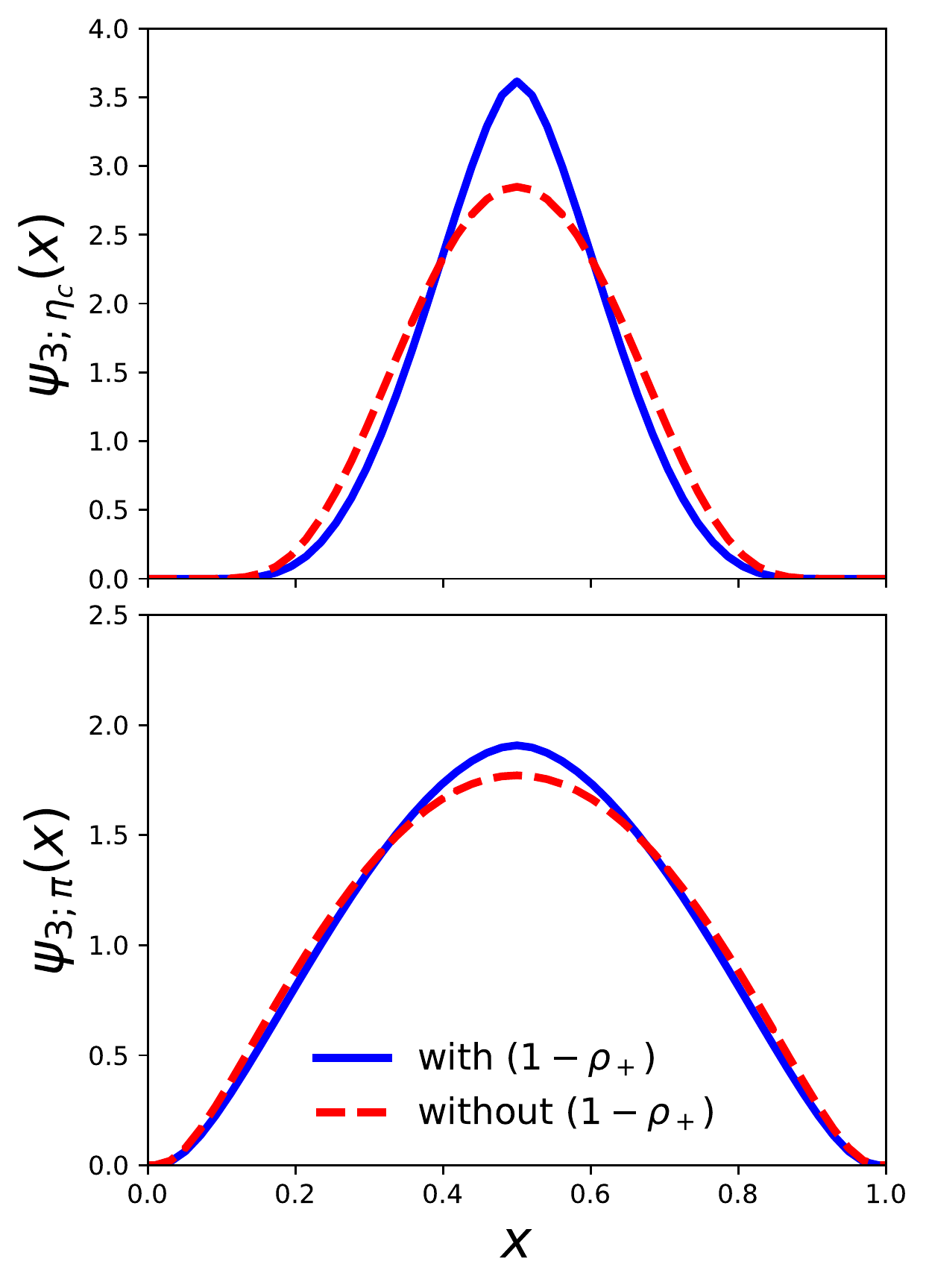}
	\caption{ \label{fig:comparison} Comparison of $\psi_{3;P}(x)$ with and without the inclusion of $(1-\rho_+)$ factor.}
\end{figure}

Figure~\ref{fig:da} presents the DAs of different twists for the $1S$ and $2S$ state heavy pseudoscalar mesons. 
Note here that the longitudinal momentum $x$ is carried by the lighter quark. As a result, the DAs for $D_{(s)}$ and $B_{(s)}$ are more concentrated in the lower $x$ region.
For the equal mass case such as $\eta_c$ and $\eta_b$, the $\phi_{2;{\rm P}}(x),$ $\phi_{3;{\rm P}}(x)$ and $\phi_{4;{\rm P}}(x)$ 
have the same shape since the corresponding operators are the same as shown in Table~\ref{tab:coeff}. 
The key reason for this is the utilization of the self-consistent condition for the replacement of $M$ with $M_0$ when obtaining the aforementioned results.
But, the $\psi_{3;{\rm P}}(x)$ has a different shape with a narrower and higher peak.
In addition, the distance between the peaks becomes closer for $\psi_{3;{\rm P}}(x)$ of the $2S$ state.
For the unequal-mass case such as $D$ or $B$ meson, 
the peak is shifted to the lower $x$ region for the higher twist DAs where the $\phi_{4;{\rm P}}(x)$ has the highest peak. 
For the $2S$ state, the peak near $x=0$ is more enhanced for the higher twist while the peak near $x=0.5$ is suppressed and shifted to the lower $x$ region.
Furthermore, the dip between the peaks is enhanced for the higher twist.

In order to gain a deeper understanding of the structure of DAs, we construct a 2D plot illustrating 
the DAs, $\phi(x)\equiv\{\phi_{n;{\rm P}}, \psi_{3;{\rm P}}\}$, using the following definition:
\begin{eqnarray}
  \phi(x) = \int^\infty_0 {\rm d}^2{\bf k}_\perp\psi(x, {\bf k}_\perp) = \int_0^1 {\rm d}y\ \phi(x,y),
\end{eqnarray}
where the wave function $\phi(x,y)=\pi\psi(x,y)/(1-y)^2$ is obtained by using the variable transformation ${\bf k}_\perp^2 = y/(1-y)$ so that $y$ ranges from 0 to 1.
For the sake of demonstration, we show the 2D plot of $\phi(x,y)$ only for $D(2S)$ as shown in Fig.~\ref{fig:da_2d}.
One can clearly see that the $\phi(x,y)$ bears resemblance with the LFWF shown in Fig.~\ref{fig:wavefunction1}(a). 
The two-peak structure in the DAs of the $2S$ state clearly originated from the nodal structure shown as the white bands. 
It appears that the $\phi_{2;{\rm P}}(x,y)$, $\phi_{3;{\rm P}}(x,y)$, and $\phi_{4;{\rm P}}(x,y)$ have a similar shape.
But, the DAs with the higher twist are more concentrated in the lower $x$ region.
On the other hand, $\psi_{3;{\rm P}}(x,y)$ has a smaller nodal structure so that it shows slightly different behavior in Fig.~\ref{fig:da}.

Figure~\ref{fig:da_hel} shows the helicity contributions to DAs of $\eta_c(1S)$ for various twist.
The dashed and dotted lines represent the ordinary $H_{\uparrow\downarrow} + H_{\downarrow\uparrow}$ and higher $H_{\uparrow\uparrow} + H_{\downarrow\downarrow}$ helicity contributions, respectively. The solid lines represent the full results.
As mentioned earlier, the DAs of twist 2, twist 3, and twist 4 for equal-mass cases are the same.
However, these DAs exhibit distinct helicity contributions. 
Specifically, the twist-2 DAs are exclusively composed of the ordinary helicity component, 
while the higher twist DAs incorporate a finite contribution from the higher helicity component.
This observation is in accordance with the result presented for helicity contribution to the decay constant in Fig.~\ref{fig:helicity}.

Although the helicity contribution to DAs generally remains unchanged regardless of whether the light quark is assigned to $x$ or $(1-x)$, 
it is crucial to acknowledge that the specific choice between $x$ and $1-x$ does impact the helicity contribution to the DA $\psi_{3;{\rm P}}(x)$ obtained from the nonlocal matrix element.
In particular, when assigning the light quark to either $x$ or $(1-x)$, the helicity contribution to $\psi_{3;{\rm P}}(x)$ exhibits markedly distinct behaviors.
Figure~\ref{fig:da3t_hel} illustrates, as an example, the discrepancy in helicity contributions to $\psi_{3;{\rm P}}(x)$ for $D(1S)$ when the 
light quark is assigned to carry either $x$ or $(1-x)$.
The upper and lower panels in Fig.~\ref{fig:da3t_hel} represent the results obtained when the light quark is assigned to $x$ and $1-x$, respectively,
and the same line codes are used as in Fig.~\ref{fig:da_hel}.
One can clearly see from Fig.~\ref{fig:da3t_hel}
that the ordinary helicity exhibits a negative contribution when the light quark carries the value of $x$ (upper panel), 
while it yields a positive contribution when the heavy quark carries the value of $x$ (lower panel).
This distinct behavior observed can be attributed to the integration over $x'$, where the behavior depends on the specific choice of $x$. It indicates that each individual ordinary and higher helicity contribution 
to $\psi_{3;{\rm P}}(x)$ depends on whether the light quark or the heavy quark carries the specific light-front longitudinal momentum fraction $x$. However, it is important to note that the total DA remains unchanged whether we assign $x$ to the light or heavy quark.
Furthermore, it is worth noting that there is a substantial cancellation between the ordinary and higher helicity contributions when $x$ is associated with the light quark, whereas the cancellation is comparatively smaller when $x$ is assigned to the heavy quark.

\begin{table}[t]
	\begin{ruledtabular}
		\renewcommand{\arraystretch}{1.3}
		\caption{ \label{tab:xi_moment1S}
		The $\xi$-moment up to $n=6$ for the $1S$ state heavy pseudoscalar mesons. Here we define the $x$ carried by the lighter quark. Therefore, the odd-power of $\expval{\xi}$ has an opposite sign to our previous work~\cite{Arifi22}. 
        }
		\begin{tabular}{cc|ccccccc}
			$(1S)$	&	tw	&$D$ & $D_s$ & $\eta_c$ & $B$ & $B_s$ & $B_c$ & $\eta_b$  \\ \hline
			$\expval{\xi^1}$ 
                    & 2  & $-0.337$ & $-0.294$ & \dots & $-0.644$ & $-0.614$ & $-0.390$ & \dots \\
                    & 3p & $-0.445$ & $-0.365$ & \dots & $-0.713$ & $-0.670$ & $-0.419$ & \dots \\
                    & 4  & $-0.553$ & $-0.436$ & \dots & $-0.781$ & $-0.726$ & $-0.447$ & \dots \\ 
                    & 3t & $-0.445$ & $-0.365$ & \dots & $-0.713$ & $-0.670$ & $-0.417$ & \dots\\ \hline
        	$\expval{\xi^2}$ 
                    & 2  & 0.226 & 0.197 & 0.088 & 0.453 & 0.417 & 0.201 & 0.049 \\
                    & 3p & 0.312 & 0.242 & 0.088 & 0.545 & 0.487 & 0.223 & 0.049 \\
                    & 4  & 0.397 & 0.288 & 0.088 & 0.636 & 0.558 & 0.246 & 0.049 \\ 
                    & 3t & 0.273 & 0.202 & 0.053 & 0.533 & 0.475 & 0.205 & 0.029\\\hline
        	$\expval{\xi^3}$ 
                    & 2  & $-0.145$ & $-0.114$ & \dots & $-0.337$ & $-0.302$ & $-0.113$ & \dots \\
                    & 3p & $-0.222$ & $-0.154$ & \dots & $-0.435$ & $-0.373$ & $-0.129$ & \dots\\
                    & 4  & $-0.299$ & $-0.193$ & \dots & $-0.534$ & $-0.445$ & $-0.146$ & \dots\\ 
                    & 3t & $-0.177$ & $-0.116$ & \dots & $-0.413$ & $-0.350$ & $-0.108$ & \dots\\ \hline
        	$\expval{\xi^4}$ 
                    & 2  & 0.108 & 0.083 & 0.018 & 0.261 & 0.228 & 0.068 & 0.006 \\
                    & 3p & 0.173 & 0.112 & 0.018 & 0.360 & 0.296 & 0.080 & 0.006 \\
                    & 4  & 0.238 & 0.142 & 0.018 & 0.458 & 0.364 & 0.092 & 0.006 \\ 
                    & 3t & 0.124 & 0.072 & 0.008 & 0.329 & 0.265 & 0.061 & 0.003\\ \hline
        	$\expval{\xi^5}$ 
                    & 2  & $-0.082$ & $-0.058$ & \dots & $-0.209$ & $-0.178$ & $-0.043$ & \dots\\
                    & 3p & $-0.138$ & $-0.082$ & \dots & $-0.304$ & $-0.241$ & $-0.052$ & \dots\\
                    & 4  & $-0.195$ & $-0.106$ & \dots & $-0.399$ & $-0.304$ & $-0.060$ & \dots \\ 
                    & 3t & $-0.090$ & $-0.048$ & \dots & $-0.267$ & $-0.206$ & $-0.035$ & \dots\\ \hline
        	$\expval{\xi^6}$ 
                    & 2  & 0.065 & 0.044 & 0.005 & 0.172 & 0.143 & 0.029 & 0.001 \\ 
                    & 3p & 0.115 & 0.063 & 0.005 & 0.262 & 0.200 & 0.035 & 0.001 \\ 
                    & 4  & 0.165 & 0.083 & 0.005 & 0.352 & 0.258 & 0.041 & 0.001 \\ 
                    & 3t & 0.068 & 0.033 & 0.002 & 0.220 & 0.162 & 0.021 & 0.0004 \\
        \end{tabular}
		\renewcommand{\arraystretch}{1}
	\end{ruledtabular}
\end{table}

\begin{table}[t]
	\begin{ruledtabular}
		\renewcommand{\arraystretch}{1.3}
		\caption{ \label{tab:xi_moment2S}
		The $\xi$-moment up to $n=6$ for the $2S$ state heavy pseudoscalar mesons. Here we define the $x$ carried by the lighter quark. Therefore, the odd-power of $\expval{\xi}$ has an opposite sign to our previous work~\cite{Arifi22}.  
        }
		\begin{tabular}{cc|ccccccc}
			$(2S)$	&	tw	&$D$ & $D_s$ & $\eta_c$ & $B$ & $B_s$ & $B_c$ & $\eta_b$  \\ \hline
			$\expval{\xi^1}$ 
                    & 2  & 0.042    & $-0.015$ & \dots & $-0.426$ & $-0.411$ & $-0.275$ & \dots \\
                    & 3p & $-0.161$ & $-0.155$ & \dots & $-0.549$ & $-0.518$ & $-0.333$ & \dots \\
                    & 4  & $-0.363$ & $-0.296$ & \dots & $-0.672$ & $-0.625$ & $-0.390$ & \dots \\ 
                    & 3t & $-0.161$ & $-0.155$ & \dots & $-0.549$ & $-0.518$ & $-0.333$ & \dots\\ \hline 
        	$\expval{\xi^2}$ 
                    & 2  & 0.052 & 0.132 & 0.179 & 0.198 & 0.202 & 0.160 & 0.099 \\
                    & 3p & 0.164 & 0.199 & 0.179 & 0.344 & 0.322 & 0.201 & 0.099 \\
                    & 4  & 0.275 & 0.266 & 0.179 & 0.491 & 0.442 & 0.241 & 0.099 \\ 
                    & 3t & 0.081 & 0.115 & 0.107 & 0.320 & 0.295 & 0.163 & 0.059\\ \hline
        	$\expval{\xi^3}$ 
                    & 2  & 0.004    & $-0.055$ & \dots & $-0.094$ & $-0.112$ & $-0.101$ & \dots \\
                    & 3p & $-0.099$ & $-0.121$ & \dots & $-0.237$ & $-0.225$ & $-0.131$ & \dots\\
                    & 4  & $-0.202$ & $-0.187$ & \dots & $-0.380$ & $-0.337$ & $-0.161$ & \dots\\ 
                    & 3t & $-0.029$ & $-0.061$ & \dots & $-0.194$ & $-0.180$ & $-0.092$ & \dots\\ \hline
        	$\expval{\xi^4}$ 
                    & 2  & 0.012 & 0.065 & 0.048 & 0.043 & 0.070 & 0.071 & 0.016 \\
                    & 3p & 0.091 & 0.113 & 0.048 & 0.175 & 0.171 & 0.093 & 0.016 \\
                    & 4  & 0.171 & 0.161 & 0.048 & 0.307 & 0.272 & 0.115 & 0.016 \\ 
                    & 3t & 0.018 & 0.048 & 0.021 & 0.120 & 0.116 & 0.057 & 0.007\\ \hline
        	$\expval{\xi^5}$ 
                    & 2  & $-0.006$ & $-0.047$ & \dots & $-0.017$ & $-0.048$ & $-0.051$ & \dots\\
                    & 3p & $-0.076$ & $-0.089$ & \dots & $-0.137$ & $-0.138$ & $-0.067$ & \dots\\
                    & 4  & $-0.146$ & $-0.131$ & \dots & $-0.257$ & $-0.228$ & $-0.084$ & \dots \\ 
                    & 3t & $-0.011$ & $-0.035$ & \dots & $-0.075$ & $-0.078$ & $-0.037$ & \dots\\ \hline
        	$\expval{\xi^6}$ 
                    & 2  & 0.010 & 0.044 & 0.016 & 0.004 & 0.036 & 0.038 & 0.003 \\
                    & 3p & 0.070 & 0.079 & 0.016 & 0.112 & 0.116 & 0.050 & 0.003 \\ 
                    & 4  & 0.129 & 0.113 & 0.016 & 0.220 & 0.196 & 0.063 & 0.003 \\
                    & 3t & 0.009 & 0.028 & 0.005 & 0.047 & 0.056 & 0.025 & 0.001 \\
        \end{tabular}
		\renewcommand{\arraystretch}{1}
	\end{ruledtabular}
\end{table}

As we previously explained in the definition of $\psi_{3;{\rm P}}(x)$ provided by Eq.\eqref{eq:tensor}, there exist two variations 
in the QCD sum-rules for defining $\psi_{3;{\rm P}}(x)$, namely, with the inclusion of $(1-\rho_+)$\cite{Ball90} or without it~\cite{Ball:2006wn}.
In the previous work~\cite{Choi:2017uos}, the $\psi_{3;{\rm P}}(x)$ DA was computed without the term $(1-\rho_+)$.
However, we found that the inclusion of $(1-\rho_+)$ is pivotal in obtaining the identical decay constants from the nonlocal pseudotensor channel as those derived from the local axial-vector and pseudoscalar channels.

Figure~\ref{fig:comparison} depicts a comparison of $\psi_{3;P}(x)$ obtained with the term $(1-\rho_+)$ (solid lines)
and without it (dashed lines), for the cases of heavy $\eta_c(1S)$ (upper panel) and the light $\pi(1S)$ (lower panel) mesons.
We note that the same model parameters are used as in~\cite{Choi:2017uos} for the plots of $\pi$.
The analysis reveals that the inclusion of the term $(1-\rho_+)$ in $\psi_{3;{\rm P}}(x)$ leads to the narrower and higher shape compared to the case where $(1-\rho_+)$ is absent.
The quantitative impact of $(1 -\rho_+)$ on $\psi_{3;P}(x)$ is found to be more significant in the heavy quark sector compared to the light quark sector.

Finally,
we also compute the $\xi$-moment up to $n=6$ defined by
\begin{eqnarray}
    \expval{\xi^n} = \int_0^1 {\rm d}x\ \xi^n\ \phi(x), 
\end{eqnarray}
where $\xi = x - (1-x) = 2x -1$.
The results are shown in Tables~\ref{tab:xi_moment1S} and \ref{tab:xi_moment2S} for the $1S$ and $2S$ state heavy pseudoscalar mesons, respectively.
Here, we note that $x$ is carried by the light quark, so the sign is opposite for the odd $\xi$-moment as compared to our previous work~\cite{Arifi22}.
For the unequal-mass case, we observe that the absolute value of odd $\xi$-moment is getting larger for the higher twist, indicating the DAs have more deviated from the center. 
The minus sign shows that the DAs are shifted to the lower $x$ region. 
We also note that the absolute value of the even $\xi$ moments increases for the higher twist. 
The absolute values of $\xi$ moments for the $\psi_{3;{\rm P}}(x)$ are generally smaller than those of $\phi_{3;{\rm P}}(x)$.

\section{Summary} 

We have investigated the decay constants and the DAs up to the twist-4 
for the $1S$ and $2S$ state heavy pseudoscalar mesons in the LFQM, 
computing the local and nonlocal matrix elements $\la 0|{\bar q}{\Gamma} q|P\ra$
with three different current operators ${\Gamma}=(\gamma^\mu\gamma_5, i\gamma_5,\sigma^{\mu\nu}\gamma_5)$. 

In our LFQM, we performed a comprehensive analysis utilizing a general reference frame where ${\bf P}_\perp\neq 0$ and explored all possible 
components of the currents. Our explicit results demonstrate the equality of the three pseudoscalar meson decay constants derived from the three distinct 
current operators $\Gamma$. This remarkable consistency in decay constants is achieved when we enforce the self-consistency condition,
i.e. the replacement of the physical mass $M$ with the invariant mass $M_0$, within 
the LFQM. This condition stems from the Bakamjian-Thomas (BT) construction, in which the meson state is based on a noninteracting quark-antiquark representation.
It is important to note that the inclusion of the $(1-\rho_+)$ factor in the definition of the nonlocal matrix elements $\la 0|{\bar q}(z)\sigma^{\mu\nu}\gamma_5 q(-z)|P\ra$ is crucial in order to obtain the same decay constant as those derived from the axial-vector and pseudoscalar currents.
In addition to secure the process-independent pseudoscalar meson decay constant, regardless of the choice of current operators $\Gamma$, we also demonstrated its explicit Lorentz and rotation invariance. 

We also examined the helicity contributions to the decay constants, offering additional insights into the structural aspects of the decay constant.
While the decay constant remains unchanged regardless of the choice of currents, the helicity contributions to the decay constant exhibit variations 
depending on the specific current and its components, as illustrated in Table~\ref{tab:coeff2}. 
%The helicity contributions provide more insight to the structure of decay constant.
As illustrated in Fig.~\ref{fig:helicity}, while the good (plus) current only receives the ordinary helicity contributions ($\uparrow\downarrow,\downarrow\uparrow$), 
the other components including the bad (minus) current 
receive the higher helicity contributions $(\uparrow\uparrow,\downarrow\downarrow)$.
We further explored the impact of $\mathbf{P}_\perp$ dependence on the helicity contributions when considering the axial-vector current with the minus current component. Notably, it becomes evident that the higher helicity contributions play a more prominent role in the low $\mathbf{P}_\perp$ region, particularly for the $2S$ state. 
These observations are depicted in Fig.~\ref{fig:pt}.

According to the classification provided in Table~\ref{tab:twist}, employing various current operators and different components of the currents leads to distinct twists in the DAs. In particular, we explored
the twist-4 DA derived from the minus component of the axial-vector current.

The various twist DAs for the $1S$ and $2S$ heavy pseudoscalar mesons are exhibited in Fig.~\ref{fig:da}.
It is evident that the higher twist DAs for the unequal-mass case is more concentrated in the lower $x$ region. 
The $\xi$ moments are also computed for the various twist DAs. 
One of the notable results is that the odd $\xi$-moment is getting larger for the higher twist.

We expect that our result is useful for the calculation of hard exclusive processes based on the QCD factorization. 
Especially, the  higher twist DAs may be important in the low-$Q^2$ region~\cite{Ryu:2018egt,Raha:2010kz}.
It would be of great importance to extend our analysis to the vector mesons with the longitudinal and transverse polarizations for further test of our methodology~\cite{Arifi:2022qnd}. 
Moreover, the investigation of the decay constant for the excited scalar, axial-vector, and tensor mesons would be also interesting to confirm whether our LFQM based on the BT construction can universally be applicable regardless of the meson quantum numbers~\cite{Chang:2018zjq, Cheng04}. 
Finally, the extension of our approach to encompass three-point functions, such as elastic or transition form factors, deserves a thorough investigation to explore the LF zero-mode effects.

\section*{Acknowledgement}

The work of A.J.A. is supported by the RIKEN special postdoctoral
researcher program and the Young Scientist Training (YST) Program at the Asia Pacific Center for Theoretical Physics (APCTP) through the Science and Technology Promotion Fund and Lottery Fund of the Korean Government and also by the Korean Local Governments -- Gyeongsangbuk-do Province and Pohang City.
The work of H.-M.C. was supported by the National Research Foundation of Korea (NRF) under Grant No. NRF- 2023R1A2C1004098.
The work of C.-R.J. was supported in part by the U.S. Department of Energy (Grant No. DE-FG02-03ER41260). 
The National Energy Research Scientific Computing Center (NERSC) supported by the Office of Science of the U.S. Department of Energy 
under Contract No. DE-AC02-05CH11231 is also acknowledged.

\appendix
\begin{table*}[t]
	\begin{ruledtabular}
		\renewcommand{\arraystretch}{1.6}
		\caption{ BS and LFQM operators of decay constant with various currents and their corresponding DA for both unequal and equal masses of the constitutes. Here we have defined 
  $\mathcal{A}=(1-x)m_1 + x m_2$, $\Delta_1=(m_1^2 +\mathbf{k}_\perp^2)/x$, $\Delta_2=(m_2^2 +\mathbf{k}_\perp^2)/(1-x)$, $\mu^0_M=M_0^2/(m_1+m_2)$, 
  and $\rho^0_+=(m_1+m_2)^2/M_0^2$. It is evident that the BS operator will be the same with the LFQM operator if we apply the $M\to M_0$ replacement. }
		\label{tab:coeff}
		\begin{tabular}{cccccc}
		$\mathcal{O}$ & Current & $\mathcal{O}_{\rm BS}^{\rm on}(m_1\neq m_2)$ &  $\mathcal{O}_{\rm LFQM}$($m_1\neq m_2)$  & $\mathcal{O}_{\rm BS}^{\rm on}(m)$ & $\mathcal{O}_{\rm LFQM}(m)$ \\ \hline 
			\multirow{4}{*}{$\mathcal{O}_{\rm A}^{\mu}$}  &  \multirow{2}{*}{$\Gamma_{\rm A}^+,\Gamma_{\rm A}^\perp$} & \multirow{2}{*}{$2\mathcal{A}$} &  \multirow{2}{*}{$2\mathcal{A}$} & \multirow{2}{*}{$2m$} & \multirow{2}{*}{$2m$} \\
			& & &   \\ 
			& \multirow{2}{*}{$\Gamma_{\rm A}^-$}   &  \multirow{2}{*}{$\dfrac{2(m_1\Delta_2+m_2\Delta_1+\mathcal{A}\mathbf{P}_\perp^2)}{ (M^2 + \mathbf{P}_\perp^2)}$ }  &  \multirow{2}{*}{$\dfrac{2(m_1\Delta_2+m_2\Delta_1+\mathcal{A}\mathbf{P}_\perp^2)}{ (M_0^2 + \mathbf{P}_\perp^2)}$ } & \multirow{2}{*}{$2m \dfrac{(M_0^2+\mathbf{P}_\perp^2)}{(M^2+\mathbf{P}_\perp^2)}$} & \multirow{2}{*}{$2m$} \\ 
			   & & &  \\ \hline 
		\multirow{2}{*}{$\mathcal{O}_{\rm P}$}   & \multirow{2}{*}{$\Gamma_{\rm P}$} 	
            & \multirow{2}{*}{ $\dfrac{{\tilde M_0}^2}{\mu_M}$ } 
            & \multirow{2}{*}{$\dfrac{{\tilde M_0}^2}{\mu^0_M}$ } & \multirow{2}{*}{$2m\dfrac{M_0^2}{M^2}$} & \multirow{2}{*}{$2m$} \\ 
			   & & &  \\ \hline
		\multirow{2}{*}{$\mathcal{O}_{\rm T}^{\mu\nu}$}  & \multirow{2}{*}{$\Gamma_{\rm T}^{+-}$, $\Gamma_{\rm T}^{\perp-}$} 
            & \multirow{2}{*}{ $\dfrac{-12M'_0k'_z}{\mu_M (1-\rho_+)}$ }  
            &   \multirow{2}{*}{ $\dfrac{-12M'_0k'_z }{\mu^{\prime 0}_M (1-\rho^{\prime 0}_+)}$ } 
           &  \multirow{2}{*}{ $\dfrac{12m(1-2x'){M'}_0^2}{M^2 - 4m^2} $ } 
        &  \multirow{2}{*}{$\dfrac{12m(1-2x'){M}^{\prime 2}_0}{{M}^{\prime 2}_0 - 4m^2} $} \\
			   & & & \\ 
		\end{tabular}
		\renewcommand{\arraystretch}{1}
	\end{ruledtabular}
\end{table*}

\section{Link between the Covariant BS Model and the LFQM}
\label{app:bs_mod}
As we explained in the Introduction, our self-consistent LFQM results, 
e.g. Eqs.~\eqref{eq:operator} and~\eqref{eq:optensor} in this work, can also be obtained from the ``Type II" link between the
manifestly covariant BS model and the LFQM, which was first introduced in~\cite{C13}.
This is another approach to arrive at the self-consistent LFQM.
Since the detailed analysis for the link between the manifestly covariant BS model and the LFQM has already been made in the previous
works~\cite{C15,Choi:2017uos}, we shall briefly discuss the essential feature of the ``Type II" link starting from the covariant BS
model in this Appendix. 

The matrix element $A_{\rm A(P)}\equiv\bra{0} \bar{q}\Gamma_{\rm A(P)} q \ket{P}$ for the local operators 
$\Gamma_{\rm A(P)}$ in the covariant BS model is given in the one-loop approximation as
\be\label{eq:one-loop}
    A_{\rm A(P)} =i N_c \int \frac{{\rm d}^4k}{(2\pi)^4} \frac{H_0 S_{\rm A(P)}}{(p^2_1 - m^2_1 + i\epsilon)(p^2_2 - m^2_2 + i\epsilon)},
\ee
where $S_{\rm A(P)}= {\rm Tr}[\Gamma_{\rm A(P)} (\slashed{p}_{1} +m_1) \gamma_5 (-\slashed{p}_{2} +m_2)]$ is the trace term with
$p_1=P-k$ and $p_2=k$.
To regularize the loop, we use the usual multipole ansatz $H_0 = \frac{g}{D_\Lambda^2}$ with $D_\Lambda = p_1^2 - m_\Lambda^2 + i \epsilon$, where
$m_\Lambda$ plays the role of the momentum cut-off.

In this exactly solvable manifestly covariant BS model, the decay constants for the axial-vector and pseudoscalar currents 
can be obtained from the manifestly covariant calculation 
using the Feynman parametrization and the final results are given by
\begin{eqnarray} 
      f_{\rm A} &=& \frac{gN_c}{4\pi^2} \int_0^1 {\rm d}x \int_0^{1-x}  {\rm d}y\ \frac{(1-x-y)B_1}{C^2},\label{eq:cov1}\\
    f_{\rm P} &=& \frac{gN_c}{4\pi^2\mu_M} \int_0^1 {\rm d}x \int_0^{1-x}{\rm d}y\  \frac{(1-x-y)(B_2-2C)}{C^2}, \label{eq:cov2}\nonumber\\  
\end{eqnarray}
where $C=y(1-y)M^2 - xm_1^2 - ym_2^2 - (1-x-y)m_\Lambda^2$,
$B_1 = m_2 + (1-y) (m_1 - m_2)$, and $B_2 = y(1-y) M^2 + m_1 m_2$.
We should note, at this point, that the two pseudoscalar meson decay constants $f_{\rm A}$ and
$f_{\rm P}$ obtained in the BS model are not the same each other, e.g.
$f_{\rm A}=208$ MeV vs. $f_{\rm P}=225$ MeV for $D(1S)$ meson with the value of $m_\Lambda=1.673$ GeV. This contrasts to our LFQM in which we obtain the process-independent decay constant.

In parallel with the manifestly covariant calculation, we 
perform the LF calculation of Eq.~\eqref{eq:one-loop} 
by doing the LF energy integration $p^-_2$ picking up the on-mass shell pole $p^2_2 = p^2_{2\rm on}= m^2_2$ and
obtain
\be
    f_{\rm A(P)} = N_c \int^1_0 {\rm d}x \int \frac{{\rm d}^2 \mathbf{k}_\bot}{8\pi^3}\  
    \frac{\chi(x,\mathbf{k}_\perp)}{1-x} {\cal O}^{\rm A(P)}_{\rm BS}, \label{eq:bsLF}
\ee
where $\chi(x, \mathbf{k}_\perp) = 1/([x(M^2 -M_0^2)][x(M^2 -M_\Lambda^2)]^2)$
is the vertex function with $M^2_\Lambda = M^2_0 (m_1\to m_\Lambda)$
and ${\cal O}^{\rm A(P)}_{\rm BS}= i S_{\rm A(P)}/2\cal P_{\rm A(P)}$.

In contrast to the LFQM constrained by the on-mass shellness of the constituents, the LF calculation of the BS model
allows the off-mass shell quark propagators.
For the axial-vector current $\Gamma^\mu_{\rm A}=\gamma^\mu\gamma_5$ with the current component $\mu=(+, \perp)$, we find that
only the on-mass shell quark propagators contribute and the full result of the operator is obtained as
$[{\cal O}^{\rm A}_{\rm BS}]_{\rm full}=[{\cal O}^{\rm A}_{\rm BS}]^{+}_{\rm on}=[{\cal O}^{\rm A}_{\rm BS}]^{\perp}_{\rm on}=2{\cal A}$.
On the other hand, the minus component of the axial-vector current receives not only the instantaneous
but also the zero-mode contributions in addition to the on-mass shell contribution, i.e.
$[{\cal O}^{\rm A}_{\rm BS}]_{\rm full}=[{\cal O}^{\rm A}_{\rm BS}]^{-}_{\rm on}+[{\cal O}^{\rm A}_{\rm BS}]^{-}_{\rm inst}
+ [{\cal O}^{\rm A}_{\rm BS}]^{-}_{\rm z.m.} = 2{\cal A}$, where
\begin{eqnarray}\label{OBSminus}
\left[{\cal O}^{\rm A}_{\rm BS} \right]^{-}_{\rm on} &=& \frac{ 2(m_1 \Delta_1 + m_2 \Delta_2 + {\cal A}{\bf P}^2_\perp)}{M^2 + {\bf P}^2_\perp},
\nonumber\\
\left[{\cal O}^{\rm A}_{\rm BS}\right]^{-}_{\rm inst} &=& \frac{ 2 m_2 (M^2 - M^2_0)}{M^2 +{\bf P}^2_\perp},
\nonumber\\
\left[{\cal O}^{\rm A}_{\rm BS}\right]^{-}_{\rm z.m.} &=& \frac{ 2 (m_1 - m_2)Z_2}{M^2 +{\bf P}^2_\perp},
\end{eqnarray}
with $\Delta_j = (m^2_j + {\bf k}^2_\perp)/x_j (j=1,2)$ and $Z_2 = x(M^2 - M^2_0) + m^2_1 - m^2_2 + (1- 2x)M^2$.

For the pseudoscalar current $\Gamma_{\rm P}=i\gamma_5$, the full operator is obtained from the sum of the three nonvanishing
contributions, i.e. $[{\cal O}^{\rm P}_{\rm BS}]_{\rm full}=2[m_1 (m_2 -m_1) + x M^2_0]/\mu_M = [{\cal O}^{\rm P}_{\rm BS}]_{\rm on}+[{\cal O}^{\rm P}_{\rm BS}]_{\rm inst}
+ [{\cal O}^{\rm P}_{\rm BS}]_{\rm z.m.}$, where
\begin{eqnarray}\label{OBSps}
\left[{\cal O}^{\rm P}_{\rm BS} \right]_{\rm on} &=& \frac{{\tilde M}^2_0}{\mu_M},
\nonumber\\
\left[{\cal O}^{\rm P}_{\rm BS}\right]_{\rm inst} &=& \frac{(1-x)(M^2 - M^2_0)}{\mu_M},
\nonumber\\
\left[{\cal O}^{\rm P}_{\rm BS}\right]_{\rm z.m.} &=& -\frac{Z_2}{\mu_M},
\end{eqnarray}
 With those full operators $[{\cal O}^{\rm A(P)}_{\rm BS}]_{\rm full}$, the LF results for $f_{\rm A(P)}$ given by
Eq.~\eqref{eq:bsLF} are the same as the corresponding covariant ones given by Eq.~\eqref{eq:cov1}. 

The basic idea of Ref.~\cite{C13}
in obtaining the LFQM result given by Eq.~\eqref{eq:operator} 
from the BS model amplitude given by Eq.~\eqref{eq:bsLF} is to replace not only vertex function $\chi(x,{\bf k}_\perp)$
in Eq.~\eqref{eq:bsLF} with the Gaussian wave function $\Phi(x,{\bf k}_\perp)$ but also all the physical mass $M$ appeared in the
BS model with the invariant mass $M_0$ via the ``Type II" link between the BS model and the LFQM as we coined in~\cite{C13}:
\begin{eqnarray}\label{TypeII}
    \sqrt{2N_c} \frac{\chi(x,\mathbf{k}_\perp)}{(1-x)} &\to& \frac{\Phi(x,\mathbf{k}_\perp)}{ \sqrt{\mathcal{A}^2 + \mathbf{k}_\perp^2}}, \nonumber\\
    M &\to& M_0.
\end{eqnarray}

The first immediate action for the replacement of $M\to M_0$ in the LFQM is to remove the instantaneous contribution ($\propto M^2 - M^2_0$), which may appear in the covariant BS model but absent in the LFQM consistent with the BT construction. The only spurious effect, which may appear in the
LFQM, is the LF zero mode.
Furthermore, a crucial feature when utilizing the ``Type II" link between the BS model and the LFQM is solely to employ the on-mass shell BS operator. 
In other words, the LFQM operator, denoted as ${\cal O}\equiv{\cal O}_{\rm LFQM}$ and defined by Eq.~\eqref{eq:operator}, 
can be directly derived by substituting $\left[{\cal O}_{\rm BS} \right]_{\rm on}(M\to M_0)$.

For the axial-vector current case, the full BS operator $\left[{\cal O}^{\rm A}_{\rm BS}\right]_{\rm full}=2{\cal A}$ obtained from
$\left[{\cal O}^{\rm A}_{\rm BS}\right]^{+}_{\rm on}=\left[{\cal O}^{\rm A}_{\rm BS}\right]^{\perp}_{\rm on}=\left[{\cal O}^{\rm A}_{\rm BS}\right]^{-}_{\rm full}$
is shown exactly the same as the ${\cal O}^{\rm A}_{\rm LFQM}$. The plus and perpendicular components of the current are free from the instantaneous 
and zero-mode contributions, and they are indeed the ``good" components of the current. While the full operator derived from the minus component of the current shares the exact same form as those derived from the ``good" currents, this feature can be considered extremely rare. 
It should be noted that the full operator obtained by including the zero mode does not generally align with the full operator derived 
solely from the on-mass shell contribution, as evident in the case of the pseudoscalar current in which the full operator
$[{\cal O}^{\rm P}_{\rm BS}]_{\rm full}$ is different from the on-mass shell operator
$[{\cal O}^{\rm P}_{\rm BS}]_{\rm on}$ and one finds that $f_{\rm P}$ in Eq.~\eqref{eq:operator} does not match with
$f_{\rm A}$ if the full operator $[{\cal O}^{\rm P}_{\rm BS}]_{\rm full}$ is used for the replacement of $M\to M_0$
instead of $[{\cal O}^{\rm P}_{\rm BS}]_{\rm on}$. A similar observation has also been made in the previous analysis for the vector meson
decay constant~\cite{C13}. This indicates that the zero mode found in the BS model is no longer applicable to the LFQM. 
Instead, the replacement of $M\to M_0$ for the on-mass shell operator $\left[{\cal O}_{\rm BS} \right]_{\rm on}$, 
e.g. $\left[{\cal O}^{\rm A}_{\rm BS} \right]^{-}_{\rm on}(M\to M_0)$ 
and $\left[{\cal O}^{\rm P}_{\rm BS} \right]_{\rm on}(M\to M_0)$, can be regarded as an effective
zero-mode inclusion in the LFQM.

For the pseudotensor current, the local matrix element $\bra{0} \bar{q}(0)\Gamma_{\rm T} q(0) \ket{P}$ defined by Eq.~\eqref{eq:one-loop} in
the covariant BS model is zero in the manifestly covariant calculation. Performing the LF calculation for this local matrix element, we also confirm that the full operator
$[{\cal O}^{\rm T}_{\rm BS}]_{\rm full}$ is zero if and only if we include both nonvanishing instantaneous and zero-mode contributions, i.e., 
$[{\cal O}^{\rm T}_{\rm BS}]_{\rm full}=[{\cal O}^{\rm T}_{\rm BS}]_{\rm on}+[{\cal O}^{\rm T}_{\rm BS}]_{\rm inst}
+ [{\cal O}^{\rm T}_{\rm BS}]_{\rm z.m.} = 0$. Because of this, the decay constant for the pseudotensor current needs to be defined only through the
nonlocal matrix element $\bra{0} \bar{q}(z)\Gamma_{\rm T} q(-z) \ket{P}$ defined by Eq.~\eqref{eq:tensor}.

Defining $z^\mu = \tau \eta^\mu$ using the lightlike vector $\eta=(1,0,0,-1)$ and multiplying $(P_\mu\eta_\nu - P_\nu\eta_\mu)$ on both sides of Eq.~\eqref{eq:tensor}, 
one can rewrite Eq.~\eqref{eq:tensor} as [see Ref.~\cite{Choi:2017uos} for more detailed derivation]
\bea 
&&\bra{0}{\bar q}(\tau\eta) i(\slashed{P}\slashed{\eta} - P\cdot\eta)\gamma_5 q(-\tau\eta)\ket{{\rm P}(P)} \nonumber\\ 
&& = \frac{i}{3}f_{\rm T}\tilde{\mu}_M (P\cdot\eta)^2\int^1_0 {\rm d}x e^{i\zeta\tau P\cdot\eta}\psi_{3;{\rm P}}(x),
\eea
where $\tilde{\mu}_M =\mu_M (1 -\rho_+).$\footnote{ In Ref.~\cite{Choi:2017uos}, the term $1-\rho_+$ in Eq.~\eqref{BSpsi3} is absent but the inclusion of this term 
in this work guarantees the process-independent decay constant in the LFQM.} We then obtain 
\be\label{BSpsi3}
\psi_{3;{\rm P}}(x) = - \frac{12}{f_{\rm T} {\tilde{\mu}_M}}
\int^\infty_{-\infty}  \frac{d\tau}{2\pi} \int^x_0 dx'
e^{-i\zeta'\tau(P\cdot\eta)}{\mathcalligra M}_{\rm T},
\ee
where 
${\mathcalligra M}_{\rm T}=\la 0|{\bar q}(\tau\eta) i(\slashed{P} \slashed{\eta} - P\cdot\eta)\gamma_5 q(-\tau\eta)|M(P)\ra$
is given by the following momentum integral in the same covariant BS model as Eq.~\eqref{eq:one-loop}
\be\label{Deq:4}
{\mathcalligra M}_{\rm T} = N_c
\int\frac{d^4k}{(2\pi)^4} 
\frac{e^{-i \tau (p_2-p_1)\cdot\eta} H_0} {(p^2_1 - m^2_1 + i\epsilon)(p^2_2 - m^2_2 + i\epsilon)} S_{\rm T},
\ee
with the trace term $S_{\rm T}= {\rm Tr}[i(\slashed{P} \slashed{\eta} - P\cdot\eta)\gamma_5 (\slashed{p}_{1} +m_1) \gamma_5 (-\slashed{p}_{2} +m_2)]$. 
It is worth noting that the explicit covariant 
calculation of Eq.~\eqref{BSpsi3} is challenging due to  
the nonlocal nature of the matrix element. For the LF calculation of Eq.~\eqref{BSpsi3}, 
we apply the equal LF time condition, $z^+=0$, and choose the LF gauge $A^+=0$ so that the path-ordered gauge factor becomes unity.
We should note that the valence contribution, i.e. $[S_{\rm T}]_{\rm val}(x', {\bf k}_\perp)=[S_{\rm T}]_{\rm on} + [S_{\rm T}]_{\rm inst}$, has the same form
as the one for the local current matrix element. However, the possible zero-mode contribution $[S_{\rm T}]_{\rm z.m.}$ is different from the one obtained 
in the local current case since the trace term, as well as the vertex function, should be integrated over $x'$ before 
the integration over $x$. The nonlocal nature of the matrix element introduces a discrepancy in the power counting of the singular term ($1/x$), which is
an essential procedure to identify any possible zero modes,
compared to the local current matrix element calculation. As a consequence, this discrepancy gives rise to distinct zero modes in the nonlocal current case.
The identification of the zero mode in this nonlocal matrix element calculation is not yet known. 
However, our ``Type II" link between the covariant BS model and the LFQM applying only to the
on-mass contribution works also for the nonlocal matrix element calculation regardless of the existence of the LF zero mode.

Thus, considering only the on-mass shell contribution to the trace term, we obtain from Eqs.~\eqref{BSpsi3} and~\eqref{Deq:4}
\be\label{BSpsi3LF}
\psi_{3;{\rm P}}(x) = -\frac{3N_c}{f_{\rm T}} \int^x_0 dx' \int\frac{d^2{\bf k}_\perp}{8\pi^3} 
 \frac{\chi(x',{\bf k}_\perp)}{(1-x')} \frac{[S_{\rm T}]_{\rm on}}{P^+{\tilde\mu}_M},
\ee
where $[S_{\rm T}]_{\rm on}= 4P^+ M^\prime_0 k^\prime_z$. Now, from the normalization of $\psi_{3;{\rm P}}(x)$, i.e. $\int^1_0 {\rm d}x\;\psi_{3;{\rm P}}(x)=1$,
we obtain 
\be
    f_{\rm T} = N_c \int^1_0 {\rm d}x  \int^x_0 {\rm d}x' \int\frac{d^2{\bf k}_\perp}{8\pi^3} \  
    \frac{\chi(x,\mathbf{k}_\perp)}{1-x} \left[{\cal O}^{\rm T}_{\rm BS}\right]_{\rm on}, \label{eq:bsLFT}
\ee
where $\left[{\cal O}^{\rm T}_{\rm BS}\right]_{\rm on} =-12M^\prime_0 k^\prime_z/{\tilde\mu}_M$.

Finally, applying the ``Type II" link given by Eq.~\eqref{TypeII} to Eqs.~\eqref{eq:bsLF} and~\eqref{eq:bsLFT}, we obta1in our LFQM results
for $(f_{\rm A}, f_{\rm P}, f_{\rm T})$ defined by Eqs.~\eqref{eq:slf} and~\eqref{fptensor}.
The on-mass shell BS and LFQM operators ${\cal O}_{\rm on}$ of the three decay constants with all possible current components 
are summarized in Table~\ref{tab:coeff}.


\begin{thebibliography}{100}

%\cite{Lepage:1980fj}
\bibitem{Lepage:1980fj}
G.~P.~Lepage and S.~J.~Brodsky,
Exclusive Processes in Perturbative Quantum Chromodynamics,
\href{https://doi.org/10.1103/PhysRevD.22.2157}{Phys. Rev. D \textbf{22}, 2157 (1980).}

\bibitem{ER80}
A. V. Efremov and A. V. Radyushkin, 
Factorization and asymptotic behaviour of pion form factor in QCD,
\href{https://doi.org/10.1016/0370-2693(80)90869-2}{Phys. Lett. B \textbf{94}, 245 (1980).}


%\cite{Chernyak:1983ej}
\bibitem{Chernyak:1983ej}
V.~L.~Chernyak and A.~R.~Zhitnitsky,
Asymptotic Behavior of Exclusive Processes in QCD,
\href{https://doi.org/10.1016/0370-1573(84)90126-1}{Phys. Rept. \textbf{112} 173 (1984)}.

\bibitem{BPP}
S. J. Brodsky, H.-C. Pauli, S. S. Pinsky,
Quantum chromodynamics and other field theories on the light cone,
\href{https://doi.org/10.1016/S0370-1573(97)00089-6}{Phys. Rept. \textbf{301} 299 (1998)}.


%\cite{Ball:2006wn}
\bibitem{Ball:2006wn}
P.~Ball, V.~M.~Braun and A.~Lenz,
Higher-twist distribution amplitudes of the $K$ meson in QCD,
\href{https://doi.org/10.1088/1126-6708/2006/05/004}{JHEP \textbf{05}, 004 (2006).}

%\cite{Accardi:2012qut}
\bibitem{Accardi:2012qut}
A.~Accardi, J.~L.~Albacete, M.~Anselmino, N.~Armesto, E.~C.~Aschenauer, A.~Bacchetta, D.~Boer, W.~K.~Brooks, T.~Burton and N.~B.~Chang, \textit{et al.}
Electron Ion Collider: The Next QCD Frontier: Understanding the glue that binds us all,
\href{https://doi.org/10.1140/epja/i2016-16268-9}{Eur. Phys. J. A \textbf{52}, 268 (2016)}.

%\cite{AbdulKhalek:2021gbh}
\bibitem{AbdulKhalek:2021gbh}
R.~Abdul Khalek, \textit{et al.}
Science Requirements and Detector Concepts for the Electron-Ion Collider: EIC Yellow Report,
\href{https://doi.org/10.1016/j.nuclphysa.2022.122447}{Nucl. Phys. A \textbf{1026}, 122447 (2022).}


%\cite{Braun:2022gzl}
\bibitem{Braun:2022gzl}
V.~M.~Braun, Higher Twists,
\href{https://doi.org/10.1051/epjconf/202227401012}{EPJ Web Conf. \textbf{274}, 01012 (2022)}.

\bibitem{PK18}
P. Kroll, K. Passek-Kumerički,
Twist-3 contributions to wide-angle photoproduction of pions,
\href{https://link.aps.org/doi/10.1103/PhysRevD.97.074023}{Phys. Rev. D \textbf{97}, 074023 (2018).}

\bibitem{BBNS}
M. Beneke, G. Buchalla, M. Neubert, and C. T. Sachrajda,
QCD factorization in $B\to\pi K, \pi\pi$ decays and extraction of Wolfenstein parameters,
\href{https://doi.org/10.1016/S0550-3213(01)00251-6}{Nucl. Phys. B \textbf{606}, 245 (2001).}

\bibitem{Ball90}
P. Ball,
Theoretical update of pseudoscalar meson distribution amplitudes of higher twist: the nonsinglet case,
\href{https://iopscience.iop.org/article/10.1088/1126-6708/1999/01/010}
{J. High Energy Phys. \textbf{01}, 010 (1999).}

\bibitem{HWZ04}
T. Huang, X. H. Wu, and M. Z. Zhou, 
Twist-3 distribution amplitude of the pion in QCD sum rules,
\href{https://link.aps.org/doi/10.1103/PhysRevD.70.014013}{Phys. Rev. D \textbf{70}, 014013 (2004).}

\bibitem{Agaev05}
S. S. Agaev, 
Impact of the higher twist effects on the $\gamma\gamma^*\to\pi^0$ transition form factor,
\href{https://link.aps.org/doi/10.1103/PhysRevD.72.114010}{Phys. Rev. D \textbf{72}, 114010 (2005).};
Erratum \href{https://link.aps.org/doi/10.1103/PhysRevD.73.059902}
{Phys. Rev. D \textbf{73}, 059902 (2006).}

\bibitem{BMS06}
A. P. Bakulev, S. V. Mikhailov, and N. G. Stefanis, 
Tagging the pion quark structure in QCD,
\href {https://link.aps.org/doi/10.1103/PhysRevD.73.056002}
{Phys. Rev. D \textbf{73}, 056002 (2006)}; 
QCD-based pion distribution amplitudes confronting experimental data,
\href{https://doi.org/10.1016/S0370-2693(01)00517-2}
{Phys. Lett. B \textbf{508}, 279 (2001).}

\bibitem{MPS10} S. V. Mikhailov, A. V. Pimikov, and N. G. Stefanis, 
Endpoint behavior of the pion distribution amplitude in QCD sum rules with nonlocal condensates,
\href{https://link.aps.org/doi/10.1103/PhysRevD.82.054020}
{Phys. Rev. D \textbf{82}, 054020 (2010).}

\bibitem{PPRWG}
V. Y. Petrov, M. V. Polyakov, R. Ruskov, C. Weiss, and K. Goeke, 
Pion and photon light-cone wave-functions from the instanton vacuum,
\href{https://link.aps.org/doi/10.1103/PhysRevD.59.114018} {Phys. Rev. D \textbf{59}, 114018 (1999).}

\bibitem{NK06}
S. I. Nam and H.-Ch. Kim, 
Twist-3 pion and kaon distribution amplitudes from the instanton vacuum with flavor SU(3) symmetry breaking,
\href{https://link.aps.org/doi/10.1103/PhysRevD.74.096007}{Phys. Rev. D \textbf{74}, 096007 (2006).}

\bibitem{AB02}
E. R. Arriola and W. Broniowski, 
Pion light-cone wave-function and pion distribution amplitude in the Nambu–Jona-Lasinio model,
\href{https://link.aps.org/doi/10.1103/PhysRevD.66.094016}{Phys. Rev. D \textbf{66}, 094016 (2002).}

\bibitem{PR01}
M. Praszałowicz and A. Rostworowski, 
Pion light cone wave-function in the nonlocal NJL model,
\href{https://link.aps.org/doi/10.1103/PhysRevD.64.074003}
{Phys. Rev. D \textbf{64}, 074003 (2001).}.

\bibitem{CRS13}
L. Chang, C. D. Roberts, and S. M. Schmidt, 
Light front distribution of the chiral condensate,
\href{https://doi.org/10.1016/j.physletb.2013.09.040}{Phys. Lett. B \textbf{727}, 255 (2013).}

\bibitem{SCCRSZ}
C. Shi, C. Chen, L. Chang, C. D. Roberts, S. M. Schmidt, and H.-S. Zong, 
Kaon and pion parton distribution amplitudes to twist three,
\href{https://link.aps.org/doi/10.1103/PhysRevD.92.014035}
{Phys. Rev. D \textbf{92}, 014035 (2015).}

\bibitem{Hwang10}
C.-W. Hwang, 
Analyses of decay constants and light-cone distribution amplitudes for $s$-wave heavy meson,
\href{https://link.aps.org/doi/10.1103/PhysRevD.81.114024}
{Phys. Rev. D \textbf{81}, 114024 (2010).}

\bibitem{CJ07}
H.-M. Choi and C.-R. Ji, 
Distribution amplitudes and decay constants for $(\pi, K,\rho, K^*)$
mesons in the light-front quark model,
\href{https://link.aps.org/doi/10.1103/PhysRevD.75.034019}
{Phys. Rev. D \textbf{75}, 034019 (2007).}

%\cite{C13}
\bibitem{C13}
H.~M.~Choi and C.~R.~Ji,
Self-consistent covariant description of vector meson decay constants and chirality-even quark-antiquark distribution 
amplitudes up to twist-3 in the light-front quark model,
\href{https://doi.org/10.1103/PhysRevD.89.033011}{Phys. Rev. D \textbf{89}, 033011 (2014)}.

%\cite{C13}
\bibitem{C15}
H.~M.~Choi and C.~R.~Ji,
Consistency of the light-front quark model with chiral symmetry in the pseudoscalar meson analysis,
\href{https://link.aps.org/doi/10.1103/PhysRevD.91.014018}{Phys. Rev. D \textbf{91}, 014018 (2015)}.

%\cite{Choi:2017uos}
\bibitem{Choi:2017uos}
H.~M.~Choi and C.~R.~Ji,
Two-particle twist-3 distribution amplitudes of the pion and kaon in the light-front quark model,
\href{10.1103/PhysRevD.95.056002}{Phys. Rev. D \textbf{95} 056002 (2017).}

\bibitem{CJ99} 
H.-M. Choi and C.-R. Ji, 
\newblock Mixing angles and electromagnetic properties of ground state pseudoscalar and vector meson nonets in the light-cone quark model,
\newblock \href {https://link.aps.org/doi/10.1103/PhysRevD.59.074015}
{Phys. Rev. D \textbf{59}, 074015 (1999)}.

\bibitem{CJ99B} 
H.-M. Choi and C.-R. Ji, 
\newblock Light-front quark model analysis of exclusive $0^-\to 0^-$ semileptonic heavy meson decays,
\newblock \href {https://doi.org/10.1016/S0370-2693(99)00817-5}
{Phys. Lett. B \textbf{460}, 461 (1999)}.

\bibitem{HJZR} 
H.-M. Choi, C.-R. Ji, Z. Li, and H.-Y. Ryu, 
\newblock Variational analysis of mass spectra and decay constants for ground state pseudoscalar and vector mesons in the light-front quark model,
\newblock \href {https://link.aps.org/doi/10.1103/PhysRevC.92.055203}
{Phys. Rev. C \textbf{92}, 055203 (2015)}.

\bibitem{Dhiman}
N. Dhiman, H. Dahiya, C.-R. Ji, and H.-M. Choi,
Twist-2 pseudoscalar and vector meson distribution amplitudes in light-front quark model 
with exponential-type confining potential,
\href{https://link.aps.org/doi/10.1103/PhysRevD.100.014026}
{Phys. Rev. D \textbf{100}, 014026 (2019).}

\bibitem{Arifi22}
A.~J.~Arifi, H.~M.~Choi, C.-R.~Ji and Y.~Oh,
\newblock Mixing effects on 1S and 2S state heavy mesons in the light-front quark model,
\newblock\href{https://link.aps.org/doi/10.1103/PhysRevD.106.014009}
{Phys. Rev. D \textbf{106}, 014009 (2022)}.

\bibitem{Jaus90}
W. Jaus, 
Semileptonic decays of $B$ and $D$ mesons in the light-front formalism, 
\href{https://link.aps.org/doi/10.1103/PhysRevD.41.3394}
{Phys. Rev. D \textbf{41}, 3394 (1990).}

\bibitem{Jaus91}
W. Jaus, 
Relativistic constituent-quark model of electroweak properties of light mesons, 
\href  {https://link.aps.org/doi/10.1103/PhysRevD.44.2851}
{Phys. Rev. D \textbf{44}, 2851 (1991).}

\bibitem{CCH97}
H.-Y. Cheng, C.-Y. Cheung, and C.-W. Hwang, 
Mesonic form factors and the Isgur-Wise function on the light front,
\href  {https://link.aps.org/doi/10.1103/PhysRevD.55.1559}
{Phys. Rev. D \textbf{55}, 1559 (1997).}

\bibitem{Choi07}
H.-M. Choi,
Decay constants and radiative decays of heavy mesons in light-front quark model,
\href {https://link.aps.org/doi/10.1103/PhysRevD.75.073016}
{Phys. Rev. D \textbf{75}, 073016  (2007).}

\bibitem{Mel1}
H. J. Melosh,
\newblock Quarks: Currents and constituents,
\href{https://link.aps.org/doi/10.1103/PhysRevD.9.1095}
{Phys. Rev. D \textbf{9}, 1095 (1988)}.

\bibitem{BT53}
B.~Bakamjian and L.~H. Thomas,
\newblock Relativistic particle dynamics. II,
\href{https://doi.org/10.1103/PhysRev.92.1300}
{Phys. Rev. \textbf{92}, 1300 (1953)}.

\bibitem{KP91}
B.~D. Keister and W.~N. Polyzou,
\newblock Relativistic Hamiltonian dynamics in nuclear and particle physics,
\newblock Adv. Nucl. Phys. \textbf{20}, 225 (1991).

\bibitem{Keister} 
B.D. Keister,
\newblock Rotational covariance and light-front current matrix elements,
\newblock \href {https://link.aps.org/doi/10.1103/PhysRevD.49.1500}
{Phys. Rev. D \textbf{49}, 1500 (1994)}.

%\cite{deMelo:1997cb}
\bibitem{deMelo:1997cb}
J.~P.~C.~B.~de Melo,  T. Frederico, H.W.L. Naus, P.U. Sauer,
Covariance of light-front models: pair current,
\href{https://doi.org/10.1016/S0375-9474(99)00371-1}{Nucl. Phys. A \textbf{660}, 219 (1999).}

\bibitem{CJ98} 
H.-M. Choi and C.-R. Ji, 
\newblock Nonvanishing zero modes in the light-front current,
\newblock \href {https://link.aps.org/doi/10.1103/PhysRevD.58.071901}
{Phys. Rev. D \textbf{58}, 071901(R) (1998)}.

\bibitem{DSHwang}
S. J. Brodsky, D. S. Hwang,
Exact light-cone wavefunction representation of matrix elements of electroweak currents,
\href{https://doi.org/10.1016/S0550-3213(98)00807-4}{Nucl. Phys. B \textbf{543}, 239 (1999).}

\bibitem{Jaus99} 
W. Jaus, 
\newblock Covariant analysis of the light-front quark model,
\newblock \href{https://link.aps.org/doi/10.1103/PhysRevD.60.054026}
{Phys. Rev. D \textbf{60}, 054026 (1999)}.

\bibitem{deMelo:02}
J.P.B.C. de Melo, T. Frederico, E. Pace, G. Salmè,
Pair term in the electromagnetic current within the Front-Form dynamics: spin-0 case,
\href{https://doi.org/10.1016/S0375-9474(02)00990-9}{Nucl. Phys. A \textbf{707}, 399 (2002).}


\bibitem{BCJ01}
B. L. G. Bakker, H.-M. Choi, and C.-R. Ji,
Regularizing the divergent structure of light-front currents,
\href{https://link.aps.org/doi/10.1103/PhysRevD.63.074014}{Phys. Rev. D \textbf{63}, 074014 (2001).};
The vector meson form factor analysis in light-front dynamics,
\href {https://link.aps.org/doi/10.1103/PhysRevD.65.116001} {Phys. Rev. D \textbf{65}, 116001 (2002).}

\bibitem{Kar76}
V. A. Karmanov,
The wave functions of relativistic bound systems,
\href{http://www.jetp.ras.ru/cgi-bin/dn/e_044_02_0210.pdf}
{Zh. Eksp. Teor. Fiz. \textbf{71}, 399 (1976) [transl.: JETP \textbf{44}, 210 (1976)].}

\bibitem{Carbonell98}
J. Carbonell, B. Desplanques, V. A. Karmanov, and J. F. Mathiot, 
Explicitly covariant light-front dynamics and relativistic few-body systems,
\href {https://doi.org/10.1016/S0370-1573(97)00090-2}
{Phys. Rep. \textbf{300}, 215 (1998).}


%\cite{Chang:2018zjq}
\bibitem{Chang:2018zjq}
Q.~Chang, X.~N.~Li, X.~Q.~Li, F.~Su and Y.~D.~Yang,
Self-consistency and covariance of light-front quark models: testing via $P$, $V$ and $A$ meson decay constants, and $P\to P$ weak transition form factors,
\href{10.1103/PhysRevD.98.114018}{Phys. Rev. D \textbf{98}, 114018 (2018).}

\bibitem{Chang22}
L. Chen, Y. W. Ren, L. T. Wang and Q. Chang,
Form factors of $P\to T$ transition within the light-front quark models, 
\href{https://doi.org/10.1140/epjc/s10052-022-10391-0}{Eur. Phys. J. C \textbf{82}, 451 (2022).}



%\cite{Arifi:2022qnd}
\bibitem{Arifi:2022qnd}
A.~J.~Arifi, H.~M.~Choi, C.~R.~Ji and Y.~Oh,
Independence of current components, polarization vectors, and reference frames in the light-front quark model analysis of meson decay constants,
\href{https://link.aps.org/doi/10.1103/PhysRevD.107.053003}{Phys. Rev. D \textbf{107}, 053003 (2023)}.


\bibitem{C21}
H.-M. Choi,
\newblock Self-consistent light-front quark model analysis of $B\to D l v_l$ transition form factors,
\href{https://doi.org/10.1103/PhysRevD.103.073004}
{Phys. Rev. D \textbf{103}, 073004 (2021)}.

\bibitem{Cheng04} 
H.-Y. Cheng, C.-K. Chua, and C.-W. Hwang, 
%\Journal{\PRD}{69}{074025}{2004}.
\newblock  Covariant light-front approach for $s$-wave and $p$-wave mesons: Its application to decay constants and form factors,
\newblock \href {https://link.aps.org/doi/10.1103/PhysRevD.69.074025}
{Phys. Rev. D \textbf{69}, 074025 (2004)}.

%\cite{Cheng:2004ew}
\bibitem{Cheng:2004ew}
H.~Y.~Cheng and C.~K.~Chua,
Light-front approach for Pentaquark strong decays,
\href{https://doi.org/10.1088/1126-6708/2004/11/072}{JHEP \textbf{11}, 072 (2004)}.

\bibitem{CJM0}
H.-M. Choi and C.-R. Ji,
Relations among the light-cone quark models with the invariant meson mass scheme and the model prediction of the 
$\eta-\eta'$ mixing angle,
\href{https://link.aps.org/doi/10.1103/PhysRevD.56.6010}{Phys. Rev. D \textbf{56}, 6010 (1997).}

%\bibitem{CCP88}
%P. L. Chung, F. Coester, and W. N. Polyzou, 
%Charge form factors of quark-model pions,
%\href {https://doi.org/10.1016/0370-2693(88)90995-1}
%{Phys. Lett. \textbf{B} 205, 545 (1988)}.

%\cite{Li:2021cwv}
\bibitem{Li:2021cwv}
M.~Li, Y.~Li, G.~Chen, T.~Lappi and J.~P.~Vary,
Light-front wavefunctions of mesons by design,
\href{https://doi.org/10.1140/epjc/s10052-022-10988-5}{Eur. Phys. J. C \textbf{82}, 1045 (2022).}




%\cite{Li:2017uug}
%\bibitem{Li:2017uug}
%Y.~Li, P.~Maris and J.~Vary,
%Frame dependence of form factors in light-front dynamics,
%\href{https://doi.org/10.1103/PhysRevD.97.054034}{Phys. Rev. D \textbf{97} 054034 (2018).}




%\cite{Ryu:2018egt}
\bibitem{Ryu:2018egt}
H.~Y.~Ryu, H.~M.~Choi and C.~R.~Ji,
Systematic twist expansion of $(\eta_c,\eta_b)\to\gamma^*\gamma$ transition form factors in light-front quark model,
\href{https://doi.org/10.1103/PhysRevD.98.034018}
{Phys. Rev. D \textbf{98}, 034018 (2018)}.

%\cite{Raha:2010kz}
\bibitem{Raha:2010kz}
U.~Raha and H.~Kohyama,
Space-and Time-like Electromagnetic Kaon Form Factors,
\href{https://doi.org/10.1103/PhysRevD.82.114012}{Phys. Rev. D \textbf{82}, 114012 (2010)}.




\end{thebibliography}
\end{document}